\documentclass{autart}
\input epsf

\usepackage{picins}
\usepackage[T1]{fontenc}
\usepackage[latin9]{inputenc}
\usepackage{color}
\usepackage{amsmath}
\usepackage{amssymb}
\usepackage{graphicx}
\usepackage{esint}
\usepackage{epstopdf}
\usepackage{enumerate}
\newcommand{\blue}{\color{blue}}

\usepackage{bbm}
\newcommand{\h}[1] {\vec{#1}}
\renewcommand{\t}{^{\mbox{\tiny {T}}}}

\newcommand{\eproof}{\hfill\rule{2mm}{2mm}}

\newcommand{\bstate}{\begin{state} }
\newcommand{\estate}{ \end{state}}

\newcommand{\bass}{\begin{ass} }
\newcommand{\eass}{  \end{ass}}

\newcommand{\brem}{ \begin{remark}  }
\newcommand{\erem}{ 
\end{remark} }
\newcommand{\bthm}{\begin{theorem}  }
\newcommand{\ethm}{ 
\end{theorem} }
\newcommand{\blem}{\begin{lemma}  }
\newcommand{\elem}{ 
\end{lemma} }
\newcommand{\bcorollary}{\begin{corollary}  }
\newcommand{\ecorollary}{   
\end{corollary} }
\newcommand{\bdefn}{\begin{definition}}
\newcommand{\edefn}{  
\end{definition} }
\newcommand{\bproposition}{\begin{proposition} }
\newcommand{\eproposition}{ 
\end{proposition} }
\newcommand{\bexample}{\begin{example} \rm}
\newcommand{\eexample}{  
\end{example} }
\newcommand{\proofnow}{\noindent{\bf Proof: }}

\newtheorem{theorem}{\bf Theorem}[section]
\newtheorem{ass}{\bf Assumption}[section]
\newtheorem{lemma}{\bf Lemma}[section]
\newtheorem{definition}{\bf Definition}[section]
\newtheorem{remark}{\bf Remark}[section]
\newtheorem{corollary}{\bf Corollary}[section]
\newtheorem{proposition}{\bf Proposition}[section]
\newtheorem{example}{\bf Example}[section]
\newtheorem{state}{\bf Assumption}[section]

\DeclareFontFamily{OMX}{yhex}{}
\DeclareFontShape{OMX}{yhex}{m}{n}{<->yhcmex10}{}
\DeclareSymbolFont{yhlargesymbols}{OMX}{yhex}{m}{n}
\DeclareMathAccent{\wideparen}{\mathord}{yhlargesymbols}{"F3}

\begin{document}
\begin{frontmatter}
%\title{An Input-Delay Method for Event-Triggered Control of  Nonlinear Systems}
%\title{Event-Triggered Control for Nonlinear Systems and its Backstepping Design
%}
\title{Event-triggered controllers
	based on the supremum  norm of  sampling-induced error}
\thanks[footnoteinfo]{This research was supported under The University of Hong Kong Research Committee Post-doctoral Fellow Scheme.}

\author[A]{Lijun Zhu} \ead{ljzhu@hust.edu.cn},
\author[B] {Zhiyong Chen}\ead{zhiyong.chen@newcastle.edu.au},
\author[A] {David J. Hill} \ead{dhill@eee.hku.hk},
\author[C] {Shengli Du} \ead{shenglidu@bjut.edu.cn}

\address[A]{Key Laboratory of Imaging Processing and Intelligence Control, School of Artificial Intelligence and Automation, Huazhong University of Science and Technology, Wuhan 430074, China}
\address[B]{School of Electrical Engineering and Computing, The University of Newcastle, Callaghan, NSW 2308, Australia}  % Please supply
\address[C]{College of Automation, Faculty of Information Technology, Beijing University of Technology, Beijing, 100124, China}
%\maketitle
\begin{abstract}
The paper proposes a novel event-triggered control scheme for nonlinear
systems. % based on the input-delay method. 
Specifically, the closed-loop
system is associated with a pair of auxiliary input and output. The
auxiliary output is defined as the derivative of the continuous-time
input function, while the auxiliary input is defined as the input
disturbance caused by the sampling or equivalently the integral of
the auxiliary output over the sampling period. As a result, it forms a cyclic
mapping from the input to the output via the system dynamics
and back from the output to the input via the integral. The event-triggered
law is constructed to make the mapping contractive such that the stabilization
is achieved and an easy-to-check Zeno-free condition is provided.
Within this framework, we develop a theorem for the event-triggered control
of interconnected nonlinear systems which is employed to solve the
event-triggered control for lower-triangular systems with dynamic
uncertainties. \end{abstract}

\end{frontmatter}

\section{Introduction}

The majority of modern control systems reside in microprocessors 
and need more efficient implementation in order to reduce computation
cost, save communication bandwidth and decrease energy consumption.
Sampled-data control has been developed to fulfill these tasks where
the execution of a digital controller is scheduled among sampling
instances periodically or aperiodically. As a type of aperiodic sampling,
event-triggered control suggests scheduling based on the state
and/or sampling error of the plant and  may achieve more efficient
sampling pattern than periodic sampling. Event-triggered control 
has been developed for stabilization and tracking of individual
systems, e.g., \cite{Tallapragada2013,Marchand2013,Tabuada2007,Liu2015} and
cooperative control of networked systems, e.g., \cite{Fan2013,Seyboth2013,chen2020often}. %{\color{blue}

The two-step digital emulation is a common technique for analysis
and design of sampled-data control systems especially for nonlinear
systems. 
%A continuous-time controller is first designed for a continuous-time
%system such that the control objective is fulfilled, which is then
%discretized into a sampled-data controller for digital implementation.
For periodic sampled-data control, it is an efficient tool to
explicitly compute estimates of the maximum allowable sampling period (MASP) that guarantees
asymptotic stability of sampled-data systems, e.g., {\blue \cite{Karafyllis2009,Nesic2009}}.
% with the emulated version
%of the given continuous-time controller
% under some conditions.
Emulation is also commonly adopted for the design of event-triggered
laws where the continuous-time controller has been first proposed to make sure that
the closed-loop system has the input-to-state stability (ISS) property, e.g.,
\cite{Anta2010,DePersis2011,Liu2015,Mazo2010,Liu2015a,Tabuada2007},
with the sampling error as the external input. {\blue The ISS condition was exploited in} a max-form, e.g.  \cite{Liu2015} or in an
 ISS-Lyapunov form, e.g.  \cite{Anta2010,Liu2015a,Mazo2010,Tabuada2007} for the closed-loop system with an emulated controller and the small gain conditions were proposed to ensure the stability of the event-triggered system. 
% for the closed-loop system
%and the event-triggered controllers were designed using small gain conditions. 
%%{\color{blue}, the event-triggered problem was interpreted as an stabilization problem of interconnected hybrid systems  for which each subsystem admitted  a ISS-Lyapunov  function  and  small gain condition was proposed.}
%In \cite{Anta2010,Liu2015a,Mazo2010,Tabuada2007}, the continuous-time
%controller is assumed to make the closed-loop system admit an ISS-Lyapunov
%function and the event-triggered law is designed to ensure the derivative
%of {\color{blue}the} Lyapunov function to be negative. 
In \cite{liberzon2014lyapunov}, the event-triggered technique in \cite{Tabuada2007} 
 was interpreted as a stabilization problem of interconnected hybrid systems  for which each subsystem admits an  ISS-Lyapunov  function  and a hybrid small gain condition was proposed. As an  other variant of small gain theorem,  the cyclic small gain theorem has 
been proved effective for  event-triggered control of large-scale
systems \cite{DePersis2011,Liu2015a,Liu2015}.  Despite these progresses, design of event-triggered controllers that  exclude Zeno
behavior for complex nonlinear systems, such as interconnected systems 
where only states of some subsystems are available for the feedback,  still remains a challenging
problem.
%Despite these progresses,
%explicit event-triggered controller design and  exclusion of Zeno
%behavior for complex nonlinear systems still remains a challenging
%problem. %{\color{blue}

%The event-triggered control  using state feedback are investigated in {\color{blue}e.g.,} \cite{seuret2016lq,borgers2018periodic}, whereas the output-based
%event-triggered control problems of systems are discussed 
%in {\color{blue}e.g.,} \cite{abdelrahim2018co,borgers2018riccati,khashooei2017output,dolk2017output}. 

%Recently, the input-delay approach was proposed for the controller
%design and performance analysis of linear systems \cite{teel1998note,Fridman2004,Mirkin2007}
%and nonlinear sampled-data systems \cite{Chen2014a,Fridman08a} using the emulation
%technique. 
%%{\color{blue}
%%Razumikhin-based analysis  are  efficient  methods for interconnected system with time delays such as used in \cite{teel1998connections,jankovic2001control}. Although the  results obtained by Razumikhin-based analysis may be conservative, the constructions of the Razumikhin-based conditions are relatively easy.} 
%This paper aims to extend the emulated sampled-data control approach to the
%event-triggered control of nonlinear systems. 

The contribution of
this paper is three-fold.  First, a novel event-triggered control design method
is proposed to achieve stabilization of individual and interconnected
nonlinear systems. % that  utilizes the actuation error caused by the sampling. 
Specifically, %a pair of auxiliary input and
%output is associated with the closed-loop system. 
the auxiliary output
is defined as the derivative of the continuous-time feedback input
function, and the auxiliary input is defined as the feedback input
error caused by the sampling or equivalently the integral of the auxiliary
output over the sampling period. Consequently, a closed-loop mapping is
formed from the input to the output via the system dynamics and back
from the output to the input via the integral function. The event-triggered
law is constructed to make the mapping contractive, which is analogue to a small gain condition. {\color{blue} The proposed method  guarantees that the sampling interval approaches a constant as the system is stabilized.
A similar approach was taken in \cite{khan2019}, where the state and actuation sampling errors were utilized to construct the event-triggered law, while only the actuation sampling error is used in this paper.  }
%With this formulation, the proposed event-triggered control scheme requires the continuous-time controller render the ISS property viewing  from  the actuation  error to the system state rather than that viewing from  the state sampling error to the system state, e.g.,  in \cite{Anta2010,DePersis2011,Liu2015,Mazo2010,Liu2015a,Tabuada2007}. For the latter, exceptional examples can be found, i.e., \cite{freeman1995global}, that the ISS property cannot be guaranteed by any continuous-time controller.
{\color{blue}
 In \cite{dolk2017output,abdelrahim2018co,abdelrahim2017robust,Nesic2009,Carnevale2007}, the periodic and event-triggered  sampling relies on the calculation of MASP $T$ and   requires the existence of a particular Lyapunov-like function on the hybrid systems.}
  
% a sampled-data system is   formulated as a hybrid system and an ISS-Lyapunov or dissipativity  like function is assumed to guarantee a MASP $T$, that is the estimate of the upper bound of the sampling interval required to stabilizes  the plant  for  periodic sampling. The event-triggered law is designed to wait  $T$ units of time before it considers the event-triggering condition that utilizes both
%the actuation and the measurement error.  The calculation of MASP relies on  constants in the ISS-Lyapunov or dissipativity  assumption and might  require  the knowledge of  initial conditions.  
%As also shown in \cite{Chen2014a} and demonstrated by the numerical simulation, the sampling interval that stabilizes  the plant usually depends on  the initial condition.
% In this sense, the event-triggered law in \cite{dolk2017output,abdelrahim2018co,abdelrahim2017robust,Nesic2009,Carnevale2007}  is a semi-global approach. 
%In this paper,  the ISS condition is considered  in  the max-form and the controller design is based on  different assumptions. The proposed event-triggered law that only utilizes the actuation sampling error can lead to a global stabilization. 

Second,  the  event-triggered design method is used for the stabilization of interconnected nonlinear
systems where only partial states are available for the feedback. In \cite{Liu2015a,Liu2015},  an auxiliary dynamic system was proposed to estimate the decay rate of immeasurable states as well as measurable states and used as  the dynamic threshold for the event-triggered law. The new formulation in this paper allows to take the impact of the dynamic uncertainties explicitly into the event-triggered controller design and the proposed controller is static  and {\blue are easier to design and implement in practice}. 
%Also, we identify the system structures that  the event-triggered law need not be updated even when  the dynamics of the immeasurable states is varied.

%Second,  as opposed to
%\cite{Liu2015a,Liu2015}, a novel event-triggered controller is proposed for interconnected nonlinear
%systems with dynamic uncertainties without adding an auxiliary dynamic system, and thus the 
%controller structure is neat.
%the local (global) stabilization without Zeno behavior does not rely on a local (global) Lipschitz condition on the system dynamics. 
Third, we solve
event-triggered control for lower-triangular nonlinear systems of relative degree greater than one using the proposed design method. The dynamic uncertainties are not directly manipulated by the controller and their states are not available for the feedback.
The research \cite{Liu2015a} dealt with a simpler case where the dynamic uncertainties only 
appear at first relative degree level. In order to handle the higher relative degree,
a new recursive recipe (backstepping technique) is designed to construct  the event-triggered law.
 
%  propagate ISS properties and finally design controller to render  the ISS properties from the actuation error (state sampling error as well) to aggregated  measurable states and states of dynamic uncertainties separately.  Simultaneously, the event-triggered law is constructed based on ISS gains. Such a  recursive recipe is a common technique to construct the continuous-time ISS feedback controller for event-triggered control of lower-triangular systems. 

%The rest of this paper is organized as follows. The event-triggered
%control design for individual and interconnected systems is introduced
%in Section \ref{sec:etc}. The method is applied to {\blue solve} event-triggered
%control of lower-triangular systems in Section \ref{sec:lts}. Numerical
%simulation is presented in Section \ref{sec:ns} and the paper is
%concluded in Section \ref{sec:con}.

\textbf{Notation.} Denote $\mathbb{R}^n$ the real coordinate space of $n$ dimensions.  Denote $ \mathbb{N}$ the set of non-negative integers, i.e., $\mathbb{N}=\{0,1,2,\cdots\}$.  Denote  {\blue $ \mathbb{N}^{+}$} the set of  positive integers and $\mathbb{R}^+$ the set of positive real numbers.  Let $\|x_{[t_{1},t_{2}]}\|:=\sup_{t\in[t_1,t_2]}\|x(t)\|$ for a given signal {\blue$x:\mathbb{R}\to \mathbb{R}^{n},n\in \mathbb{N}^{+}$.} The symbol $\mbox{col}(z,x)=[z\t,x\t]\t$ denotes the stacked vector by the vectors $z$ and $x$.

%A continuous function  $\alpha :[0,a)\rightarrow [0,\infty )$ is said to belong to class   $\mathcal {K}$, i.e., $\alpha\in\mathcal{K}$,   if $\alpha$
% is strictly increasing and $\alpha(0) = 0$ where $a\in\mathbb{R}^+$.  A continuous function  $\alpha$  is said to belong to class   $\mathcal {K}_\infty$, i.e., $\alpha\in\mathcal{K}_\infty$,   if $\alpha\in\mathcal{K}$, $a=\infty$ and $\lim_{s\rightarrow\infty}\alpha(s)=\infty$.
%A continuous function  $\beta :[0,a)\times [0,\infty )\rightarrow [0,\infty )$  is said to belong to class $\mathcal {KL}$, i.e., $\beta\in\mathcal{KL}$,   if for each fixed $s$,  the function $\beta (r,s)\in\mathcal{K}$ and
%for each fixed  $r$, the function  $\beta (r,s)$ is decreasing with respect to  $s$ and $ \beta (r,s)\rightarrow 0$ for $s\rightarrow \infty $.
%
\section{Main results \label{sec:etc}}

\subsection{Event-Triggered Control Method\label{sub:cd}}
\begin{figure}
\vspace{-3mm}
\centering\includegraphics[scale=0.6]{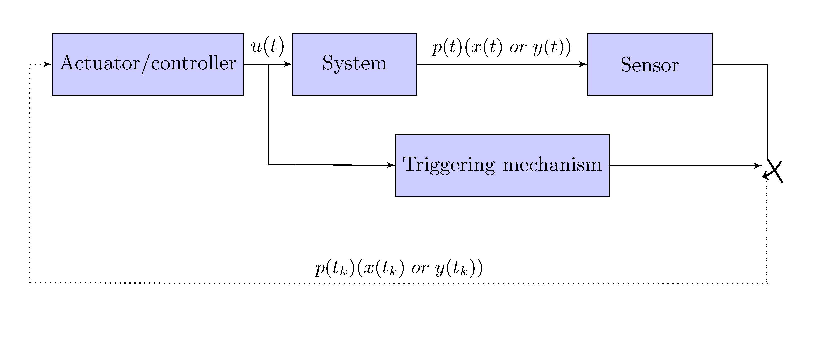}
\caption{The formulated event-triggered control system structure.\label{fig:diagram} } 
\end{figure}

Consider a nonlinear system
\begin{equation}
\dot{x}(t)=f(x(t),u(t)),\;
y(t)=h(x(t)),\label{eq:sys}
\end{equation}
where $x\in\mathbb{R}^{n}$ is the state, $y\in\mathbb{R}^{q}$ the output and $u\in\mathbb{R}^{m}$
the input. The  continuous function $f:\mathbb{R}^{n}\times\mathbb{R}^{m}\to\mathbb{R}^{n}$ satisfies
$f(0,0)=0$ such that $x=0$ is the equilibrium point of the system \eqref{eq:sys} when {\blue$u=0$}. The function $h$ is  continuously differentiable. Suppose stabilization of the equilibrium point can be
fulfilled by the continuous-time state feedback controller 
\begin{equation}
  u(t)=g(y(t)),\label{eq:con}
\end{equation}
{\color{blue}
with $g$ continuously differentiable  where it is a continuous-time state feedback controller if $y(t)=x(t)$ or otherwise a continuous-time output feedback controller.} In this paper, we
will study the event-triggered version of (\ref{eq:con}) as follows
\begin{equation}
u(t)=g( y(t_{k})),\;t\in[t_{k},t_{k+1}),\;k\in \mathbb{N}\label{eq:sample_con}
\end{equation}
where $\{t_{k}\}_{k\in \mathbb{N}}$  is a sequence of sampling time
instances and triggered by the
condition 
\begin{equation} 
t_{k+1}=\inf_{t>t_k}\{\Xi(t_k,y(t_{k}), y(t))>0\},\;k\in \mathbb{N}\label{eq:event} 
\end{equation}
{\blue with function $\Xi(t_k, y(t_{k}),y(t))$ to be designed}. 
The formulated event-triggered control system structure   is illustrated in Fig. \ref{fig:diagram}. The
objective of event-triggered control is to design the triggering law
(\ref{eq:event}) such that the closed-loop system composed of (\ref{eq:sys})
and (\ref{eq:sample_con}) achieves
\begin{enumerate}[1)]
 \item \textit{Stabilization}: the system is globally asymptotically stable
at the origin.%, i.e., $\lim_{t\rightarrow\infty}x(t)=0$; {\blue [the i.e. part is inaccurate: asymptotically stable is more than convergence ]}
\item \textit{Zeno-free Behavior}: the infinitely fast sampling is avoided, i.e., $\inf_{k\in \mathbb{N}}\{t_{k+1}-t_{k}\}>0$ for any initial conditions.
\end{enumerate}
%Except for the main Objective 1), Objective 2) guarantees that infinitely
%fast sampling is avoided. 
%The event-triggered stabilization problem
%can be solved by a small gain theorem \cite{Liu2015} if the controlled
%system has a certain input-to-state stability (ISS) property from
%the sampling-induced state error $e(t)=x(t_{k})-x(t)$  to the state $x(t)$.
%However, such an ISS property is not always achievable for complex nonlinear
%systems. An example can be found in Section III of \cite{freeman1995global}
%where for any continuous-time controller that takes the form $u(t)=g(x(t)+e(t))$ for some function $g$, there exists a solution $x(t)$ from $x(0)$
%which exhibits a finite escape time  even for
%a sufficiently small $e(t)$,  i.e., the ISS property from  $e(t)$ to the state $x(t)$ is not guaranteed.
% An alternative is to  consider  the ISS property from the auxiliary input 
%$r(t)$ defined %in (\ref{eq:r})
%later to the state $x(t)$ instead.
%We will propose a new event-triggered control that requires
%the controlled system have  %By the input-delay method,
The closed-loop system composed of (\ref{eq:sys}) and (\ref{eq:sample_con})
can be written as follows 
\begin{equation}
\dot{x}(t)  =  f(x(t),g(y(t))-r(t)),\label{eq:closed-loop}
\end{equation}
with the auxiliary input $r(t)$ and output $\xi(t)$ defined as follows
\begin{equation}
r(t)  =  \int_{t_{k}}^{t}\xi(s)ds,\;t\in[t_{k},t_{k+1}),\;
\xi(t)  = \frac{dg(y(t))}{dt}.\label{eq:r}
\end{equation}
Different from most event-triggered control designs,  the
continuous-time  feedback controller $g(y(t))$ in (\ref{eq:con}) is assumed to
ensure that the closed-loop system has the following ISS and IOS conditions. 
Since $\xi$ relies on the actuation sampling error $r$ by its definition  in (\ref{eq:r}), the IOS property characterizes the upper bound of $\|\xi(t)\|$  as  a function of $\|r_{[t_{0},t]}\|$.

\bass \label{ass:IOSS} The closed-loop system (\ref{eq:closed-loop})
with {\color{blue}$r$} as the piecewise continuous bounded external input and
{\color{blue}$\xi$} as the output has following input-to-state stability (ISS)
and input-to-output stability (IOS) properties
\begin{align}
\|x(t)\| & \leq  \max\{\tilde{\beta}(\|x(t_{0})\|,t-t_{0}),\tilde{\gamma}(\|r_{[t_{0},t]}\|)\},\label{eq:xISS}\\
\|\xi(t)\| & \leq  \max\{\beta(\|x(t_{0})\|,t-t_{0}),\gamma(\|r_{[t_{0},t]}\|)\},\label{eq:xiIOS}
\end{align}
for $ t>t_{0}$ where $\beta,\tilde{\beta}\in\mathcal{KL}$ and $\gamma,\tilde{\gamma}\in\mathcal{K}_{\infty}$.
\eass 
Then, a new event-triggered control scheme is proposed as follows.

\bthm \label{thm:etc} Consider the system (\ref{eq:sys}) with the controller
(\ref{eq:sample_con}). Suppose the closed-loop system satisfies Assumption
\ref{ass:IOSS} and the gain function $\gamma$ satisfies
\begin{equation}
\lim_{s\rightarrow0^{+}}\frac{\gamma(s)}{s}<\infty.\label{eq:glim}
\end{equation}
Let $\epsilon>1$ and $\bar{\gamma}$ be a $\mathcal{K}_{\infty}$
function satisfying 
\begin{equation}
\bar{\gamma}(s)\geq\epsilon\gamma(s),\;\forall s>0\;\mbox{and}\;\lim_{s\rightarrow0^{+}}\frac{\bar{\gamma}(s)}{s}=C_s\label{eq:gamma}
\end{equation}
for some $C_s>0$.
The objectives of the event-triggered control are achieved if the event-triggered
law (\ref{eq:event}) is 
\begin{equation}
t_{k+1}=\inf_{t>t_{k}}\{(t-t_{k})\bar{\gamma}(\|r_{[t_{k},t]}\|)>\|r_{[t_{k},t]}\|\},\;k\in \mathbb{N}\label{eq:event_law}
\end{equation}
 {\blue and Zeno-free behavior is achieved}.
\ethm

\proofnow  
The closed-loop system (\ref{eq:closed-loop}) and (\ref{eq:r}) can be regarded as
the interconnection of the $x$-subsystem and the $r$-subsystem noting that $p=x$ or $p=y$. The proof will be divided into four steps. 

%  First, we show
%that the signals $x$, $r$ and $\xi$ are bounded. Second, we prove that it is Zeno free.  Third, we  show that the  system is stable. Forth, we  show the convergence of  the system. Combining the third and forth proof, we can conclude that the closed-loop system is asymptotically stable. 

(1) Boundedness. We start with proving that $\|\xi(t)\|\leq R,\;\forall t\geq  t_0$ for any given initial condition $x(t_0)$ where  $
R:=\beta(\|x(t_{0})\|,0). $ If this is not true, there exists a finite time $T>0$ such that
$\|\xi(T)\|>R$ and $\|\xi(t)\|\leq \bar R:=R+\Delta R, \forall t_0\leq t \leq T$ for a sufficiently small $\Delta R$.  It will lead to the contradiction.
%Since $\xi(t)$ and hence $r(t)$ are bounded for $t\in[t_{0},T]$,
%using (\ref{eq:xiIOS}) and (\ref{eq:gamma}) shows that 
%\begin{equation}
%\|\xi_{[t_{0},T]}\|\leq\max\{\beta(\|x(t_{0})\|,0),\bar{\gamma}(\|r_{[t_{0},T]}\|)\}.\label{eq:xi_eq1}
%\end{equation}
%Also due to (\ref{eq:event_law}), one has  \[(t-t_{k})\bar{\gamma}(\|r_{[t_{k},t]}\|)\leq\|r_{[t_{k},t]}\|,\;  \forall t\in[t_k,t_{k+1}].\]
Denote  $\mathbb{S}_T:=\{k| k\in \mathbb{N} \mbox{ and } t_0\leq t_k\leq T$\} the set of sampling steps within $[t_0,T]$. % One has  $\inf\{t_{k+1}-t_{k}\}>0,\forall k\in\mathbb{S}_T$. 
%To see this, for each
%$k\in\mathbb{S}_T$, there exists a $t_{c}\geq0$ such that the signal
%$\|r_{[t_{k},t]}\|$ is 
%\[
%\|r_{[t_{k},t]}\|=\begin{cases}
%>0, & t>t_{k}+t_{c},\\
%=0, & t_{k}+t_{c}\geq t\geq t_{k}.
%\end{cases}
%\]
%If $t_{c}>0$, (\ref{eq:event_law}) implies $t_{k+1}-t_{k}\geq t_{c}>0$.
%If $t_{c}=0$, we will see 
%%there exists a $t_{f}$ such that $(t-t_{k})\bar{\gamma}(\|r_{[t_{k},t]}\|)<\|r_{[t_{k},t]}\|,\;\forall t_{k}< t\leq t_{f}$
%%and hence
% $t_{k+1}-t_{k}>0$ also holds.  According to (\ref{eq:glim}),
%one can always find a function $\bar{\gamma}(s)$ satisfying (\ref{eq:gamma}). Let us choose $\delta>0$.
Due to $C_s=\lim_{s\rightarrow0^{+}}\bar{\gamma}(s)/s$, for a given  $\delta$, there exists an $\epsilon$ such
that 
$
0<s<\epsilon\implies\left|\bar{\gamma}(s)/s-C_s\right|<\delta.
$
Since $\xi(t) \leq \bar R, \forall t\in[t_0, T]$,  one has $\|r_{[t_{k},t]}\|< \epsilon$  for any  $t<t_k + \epsilon/ \bar R$ due to the definition of $r(t)$ in (\ref{eq:r}).  As a result, for $t<t_k+\epsilon/\bar R$, one has
\begin{eqnarray}
\|r_{[t_{k},t]}\|<\epsilon\implies\left|\frac{\bar{\gamma}(\|r_{[t_{k},t]}\|)}{\|r_{[t_{k},t]}\|}-C_s\right|<\delta.\nonumber \label{eq:ine_r}
\end{eqnarray}
Consequently, the inequality 
%(\ref{eq:ine_r})
 leads to $\mu\bar{\gamma}(\|r_{[t_{k},t]}\|)<\|r_{[t_{k},t]}\|,\;\forall t < \epsilon/\bar R+t_k$ where $\mu=1/(C_s+\delta)$. Due to the event-triggered law (\ref{eq:event_law}), $t_{k+1}-t_k\geq \min\{1/(C_s+\delta),\epsilon/ \bar R\}> 0$. So, there exist finite sampling steps within the time duration $[t_0,T]$. 
Due to $r(t)=\int_{t_{k}}^{t}\xi(s)ds,\;t\in[t_{k},t_{k+1}),$
\begin{equation}
\|r_{[t_{k},t]}\|\leq(t-t_{k})\|\xi_{[t_{k},t]}\|,\;\forall t\in[t_{k},t_{k+1})\label{eq:r_xi}
\end{equation} 
Using inequality (\ref{eq:r_xi})  and (\ref{eq:xiIOS}) leads to 
\begin{align*}
\|\xi_{[t_{0},T]}\| & \leq  \max\{R,\bar{\gamma}(\|r_{[t_{0},t_{1}]}\|),\bar{\gamma}(\|r_{[t_{1},t_{2}]}\|),\cdots,\bar{\gamma}(\|r_{[t_{i},T]}\|)\}\\
 & \leq  \max\{R,\frac{1}{t_{1}-t_{0}}\|r_{[t_{0},t_{1}]}\|,\frac{1}{t_{2}-t_{1}}\|r_{[t_{1},t_{2}]}\|,\cdots,\\
 &\quad \frac{1}{T-t_{i}}\|r_{[t_{i},T]}\| \} \leq  R
\end{align*}
which leads to a contradiction against $\xi(T)>R$. So, $\xi(t)$ is bounded,
i.e., $\|\xi_{[t_{0},\infty)}\|\leq R$.  
It follows from (\ref{eq:r_xi}) and (\ref{eq:event_law}) that
\begin{equation}
\bar{\gamma}(\|r_{[t_{k},t]}\|)\leq\|\xi_{[t_{k},t]}\|,\forall t\in[t_{k},t_{k+1}).\label{eq:ineq_fact-1}
\end{equation}
 Thus, $r(t)$ and hence $x(t)$ are bounded for $t>t_{0}$ due to
(\ref{eq:xISS}), i.e., 
\begin{align}
 \|x_{[t_{0},\infty)}\| &\leq  \max\{\tilde{\beta}(\|x(t_{0})\|,0),\tilde{\gamma}(\bar\gamma^{-1}(\beta(\|x(t_{0})\|,0)))\} \nonumber\\
\|r_{[t_{0},\infty)}\|&\leq \bar\gamma^{-1}(\beta(\|x(t_{0})\|,0)) \label{eq:inq_xr}
\end{align}

(2) Zeno-free behavior.  The avoidance of  Zeno behavior  follows from the proof of    $\inf_{k\in \mathbb{N}}\{t_{k+1}-t_{k}\}>0,\forall k\in\mathbb{S}_T$ in the first step.  We only need to  replace the argument that  $
\xi(t)\leq R$ is bounded for $t\in[t_0,T]$  with that  $
\xi(t)\leq R$  for $t>t_0$ and the set $\mathbb{S}_T$ with $\mathbb{N}$. And it also shows Zeno-free behavior can be achieved globally.

(3) Stability. The equilibrium point $x =0$ of the closed-loop system is
stable, that is,  for any $\delta> 0$, there exists an $\epsilon$ such that if $\|x(t_{0})\|\leq\epsilon$ implies $\|x_{[t_{0},\infty)}\| \leq\delta$. From (\ref{eq:inq_xr}), it suffices to choose  $\epsilon$ such that $\max\{\tilde{\beta}(\epsilon,0),\tilde{\gamma}(\bar\gamma^{-1}(\beta(\epsilon,0)))\} 
\leq\delta$ which is  feasible.

(4) Convergence.
The final step is to show that the state $x(t)$ approaches zero
asymptotically
$
\lim_{t\rightarrow\infty}x(t)=0.
$
Due to (\ref{eq:gamma}), one has 
\begin{equation}
\inf_{s\in(0,r_{\infty})}\frac{\bar{\gamma}(s)}{s}\geq1/T_{\max}\label{eq:tmax1}
\end{equation}
for some constant $T_{\max}<\infty$, which implies that $T_{\max}\bar{\gamma}(\|r_{[t_{k},t_{k}+T_{\max}]}\|)\geq\|r_{[t_{k},t+T_{\max}]}\|$
and thus 
\begin{equation}
t_{k+1}-t_{k}\leq T_{\max},\;\forall t\in \mathbb{N},\label{eq:tmax}
\end{equation}
i.e., $T_{\max}$ is the upper bound of the sampling interval. Consider
the system behaviors of $\xi$ among interval $[t^{*}/2,t^{*}]$ for
any $t^{*}>8T_{\max}$. First, inequality (\ref{eq:xiIOS}) with
$t_{0}=t^{*}/4$ implies the signal $\|\xi_{[t^{*}/2,t^{*}]}\|$ satisfies
\begin{equation}
\|\xi_{[t^{*}/2,t^{*}]}\|\leq\max\{\beta(\|x(\frac{t^{*}}{4})\|,\frac{t^{*}}{4}),\gamma(\|r_{[t^{*}/4,t^{*}]}\|)\}.\label{eq:xi_b-1}
\end{equation}
Note that there exists integers $i,j$ such that $t^{*}/4\in[t_{i},t_{i+1}]$
where $t_{i}\geq t^{*}/8$, and $t^{*}\in[t_{j},t_{j+1}]$. Then,
it follows from (\ref{eq:xi_b-1}) and (\ref{eq:gamma}) that
\begin{align}
&\|\xi_{[t^{*}/2,t^{*}]}\|  \leq  \max\{\beta(\|x(\frac{t^{*}}{4})\|,\frac{t^{*}}{4}),1/\epsilon\bar{\gamma}(\|r_{[t_{i},t^{*}]}\|)\}\nonumber \\
% & \leq  \max\{\beta(\|x(\frac{t^{*}}{4})\|,\frac{t^{*}}{4}),\frac{1}{\epsilon(t_{i+1}-t_{i})}\|r_{[t_{i},t_{i+1}]}\|,\nonumber\\
% &\quad \cdots, \frac{1}{\epsilon(t^{*}-t_{j})}\|r_{[t_{j},t^{*}]}\|\}\nonumber \\
 %& \leq  \max\{\beta(\|x(\frac{t^{*}}{4})\|,\frac{t^{*}}{4}),1/\epsilon\|\xi_{[t^{*}/8,t^{*}]}\|\}\nonumber \\
 & \leq  \max\{\beta(\|x_{\infty}\|,\frac{t^{*}}{4}),1/\epsilon\|\xi_{[t^{*}/8,t^{*}/2]}\|\},\label{eq:xi_ine}
\end{align}
where the second inequality uses (\ref{eq:event_law}), the third
one uses (\ref{eq:r_xi}) and the last one uses $\|x_{[t_{0},\infty)}\|\leq x_{\infty}$.
Denote $\zeta(t^{*})=\|\xi_{[t^{*}/2,t^{*}]}\|$, (\ref{eq:xi_ine})
can be rewritten as 
\begin{equation}
\zeta(t^{*})\leq\max\{\beta(\|x_{\infty}\|,\frac{t^{*}}{4}),\frac{1}{\epsilon}\zeta(\frac{t^{*}}{2}),\frac{1}{\epsilon}\zeta(\frac{t^{*}}{4})\},\label{eq:zeta_*}
\end{equation}
for all $ t^{*}\geq8T_{\max}.$
Next, we will show that $\lim_{t^{*}\rightarrow\infty}\zeta(t^{*})=0$.
Otherwise, there exists a positive $\delta$ such that, for any $T$,
there exists $t^{\ast}>T$ such that $\zeta(t^{*})>\delta$. Pick
a positive integer $N$ satisfying
\begin{equation}
\delta\epsilon{}^{N}>R\label{eq:dR}
\end{equation}
and a $T$ such that 
$
T/4^{N}>T_{\max},\;\beta(\|x_{\infty}\|,T/4^{N})<\delta.
$
So, there exists $t^{*}>T$ such that $\zeta(t^{*})>\delta$. As a
result,
\[
\zeta(t^{*})>\delta>\beta(\|x_{\infty}\|,\frac{T}{4^{N}})>\beta(\|\Lambda_{\infty}\|,\frac{t^{*}}{4})
\]
which together with (\ref{eq:zeta_*}) implies $\zeta(t^{*})\leq1/\epsilon\zeta(\frac{t^{*}}{a_{1}}),\;\forall t^{*}\geq T$
where $a_{1}=2$ or $a_{1}=4$. By repeating this manipulation $N$
times, one has
$
\delta<\zeta(t^{*})\leq\left(\frac{1}{\epsilon}\right)^{N}\zeta(\frac{t^{*}}{a_{1}\cdots a_{N}}),\;\forall t^{*}\geq T.
$
As a result, (\ref{eq:dR}) further leads to 
$
R<\delta\epsilon{}^{N}\leq\zeta(\frac{t^{*}}{a_{1}\cdots a_{N}}),
$
which is a contradiction against $\xi(t)\leq R,\;\forall t\geq t_{0}$
proved in the second step. Consequently, the fact that $\lim_{t\rightarrow\infty}\zeta(t)=0$
holds which in turn implies $\lim_{t\rightarrow\infty}\xi(t)=0$,
$\lim_{t\rightarrow\infty}r(t)=0$, and hence $\lim_{t\rightarrow\infty}x(t)=0$. Therefore, the system is asymptotically stable. 
Thus, Objective 1) and 2) of the event-triggered control are achieved globally (as the initial condition goes to infinity).
\eproof

\brem {\blue We can always select} a function $\bar{\gamma}(s)$
that satisfies the condition (\ref{eq:gamma}) in Theorem (\ref{thm:etc}).
For instance, $\bar{\gamma}(s)=\epsilon_{1}\gamma(s)+\epsilon_{2}s$
for $\epsilon_{1}>1$ and $\epsilon_{2}>0$. \erem

%\brem The Zeno-free condition (\ref{eq:glim}) can be easily checked
%and ensured while the event-triggered law is constructed (see example
%in Section \ref{sec:ns}). It does not require locally (globally)
%Lipschitz condition on the system dynamics as normally needed in the
%event-triggered control \cite{Liu2015a,Liu2015,Tabuada2007}. \erem

\brem The first step of the proof shows that the positive lower bound of the sampling interval depends on the initial conditions. In particular, it decreases when the norm of the system initial condition increases. \erem

\brem 
The event-triggered law (\ref{eq:event_law}) leads to 
$
\frac{\bar{\gamma}(\|r_{[t_{k},t]}\|)}{\|r_{[t_{k},t]}\|}\leq\frac{1}{(t-t_{k})},\forall t\in[t_{k},t_{k+1}).
$
Note that the left hand side of the inequality is a nonlinear functional of the error $r$ and the right hand side is a time-varying threshold. It is similar to the event-triggered generated by a time-varying threshold in literature \cite{guinaldo2013distributed,seyboth2013event-based}. 
\erem
%{\color{red}
%\brem The original system has been converted into an interconnected system with time delays through the emulated sampled-data method. In \cite{teel1998connections,jankovic2001control}, some efficient Razumikhin-based analysis methods for interconnected system with time delays have been proposed. Though the constructions of the Razumikhin-based conditions are easy, the obtained results may be conservative. 
%\erem
%}

\brem Let us consider the event-triggered control law (\ref{eq:event_law}) when the system is subject to the external disturbance $w(t)$,
\begin{equation}
\dot{x}(t)=f(x(t),u(t),w(t)),\;
y(t)=h(x(t),w(t))
\end{equation}
Suppose
the  system with
 $r$  in (\ref{eq:r}) as input  and
$\xi$ in (\ref{eq:r}) as the output has  ISS 
and IOS properties, similar to Assumption \ref{ass:IOSS},
\begin{align*}
\|x(t)\| & \leq  \max\{\tilde{\beta}(\|x(t_{0})\|,t-t_{0}),\tilde{\gamma}(\|r_{[t_{0},t]}\|), \tilde{\gamma}_x(\|w_{[t_{0},t]}\|)\}\\
\|\xi(t)\| & \leq  \max\{\beta(\|x(t_{0})\|,t-t_{0}),\gamma(\|r_{[t_{0},t]}\|),\gamma_\xi(\|w_{[t_{0},t]}\|)\},
\end{align*}
for all  $t>t_{0}$ where $\beta,\tilde{\beta}\in\mathcal{KL}$ and $\gamma,\tilde{\gamma},\tilde{\gamma}_x,\gamma_\xi\in\mathcal{K}_{\infty}$.
The event-triggered control law  can achieve the following property of the closed-loop system viewing from the external disturbance $w$  to the state $x$, i.e.,
\[
\|x(t)\| \leq  \max\{\bar{\beta}(\|x(t_{0})\|,t-t_{0}), \tilde{\gamma}_w(\|w_{[t_{0},t]}\|)\},\;\forall t>t_0
\] 
for some functions  $\bar{\beta}\in\mathcal{KL}$ and $\tilde\gamma_w\in\mathcal{K}_{\infty}$. The proof is similar to that in Theorem \ref{thm:etc}. The boundedness of  system trajectories follows from the first part of the proof of Theorem \ref{thm:etc}.  Zeno-free behavior relies on the existence of the limit $\lim_{s\rightarrow0^{+}}\bar{\gamma}(s)/s$ and is not affected by the external disturbance. The proof of the  property is similar to the proof of Theorem 3.2 in \cite{chen2016robust}. 
\erem

%{\color{blue} If $\gamma(s) = s^2$, $\mu$ can be arbitrarily small so that $T$ can be arbitrarily large ???}

The following proposition shows that the sampling interval converges
to a constant and the event-triggered control tends to be periodic
sampling control as $t\rightarrow\infty$. 
%A similar property  implicitly exists in the example of \cite{liberzon2014lyapunov}.

\bproposition \label{prop:sp}Consider the event-triggered law (\ref{eq:event_law})
in Theorem \ref{thm:etc}. Suppose $\mu:=\lim_{s\rightarrow0^{+}}\bar{\gamma}(s)/s>0$
and $T=1/\mu$. Then, $\lim_{k\rightarrow\infty}t_{k+1}-t_{k}=T$ for all initial conditions.
\eproposition 

\proofnow The idea of this proof is to show that $\lim_{k\rightarrow\infty}t_{k+1}-t_{k}=T$ as $r(t)$ approaches zero when $t$ goes to infinity.  Note that it suffices to prove for any $\delta^{*}$ there exists
a $k^{*}$ such that $|t_{k+1}-t_{k}-T|<\delta^{*}$, $\forall k\geq k^{*}$.
Without loss of generality, we only consider the case of $\delta^{*}<T$.
Due to $\lim_{s\rightarrow0^{+}}\bar{\gamma}(s)/s=1/T$, for any $\delta>0$,
there exists an $\epsilon$ such that 
\begin{equation}
\|r_{[t_{k},t]}\|<\epsilon\implies\left|\frac{\bar{\gamma}(\|r_{[t_{k},t]}\|)}{\|r_{[t_{k},t]}\|}-1/T\right|<\delta.\label{eq:ine_r-1}
\end{equation}
Due to $\lim_{t\rightarrow\infty}r(t)=0$ by Theorem \ref{thm:etc},
for any $\delta_{1}>0$, there exists  $t(\delta_{1})$ such that
$t>t(\delta_{1})\implies\|r(t)\|<\delta_{1}$. In what follows, we
consider the behavior of $r(t)$ for $t>t(\delta_{1})$. Let 
\begin{equation}
0<\delta<\min\{\frac{\delta^{*}}{T^{2}+T\delta^{*}},\frac{\delta^{*}}{T^{2}-T\delta^{*}}\}.\label{eq:delta}
\end{equation}
We choose $\delta_{1}\leq\epsilon$ such that $\|r(t)\|<\epsilon$
is satisfied for $t>t(\delta_{1})$ and $k^{*}$ such that $t(k^{*})>t(\delta_{1})$.
Consequently, (\ref{eq:ine_r-1}) implies
$
T/(1+T\delta)\gamma(\|r_{[t_{k},t]}\|)<\|r_{[t_{k},t]}\|
$
and 
$
T/(1-T\delta)\gamma(\|r_{[t_{k},t]}\|)>\|r_{[t_{k},t]}\|,\;\forall k\geq k^{*},
$
which, together with the event-triggered law (\ref{eq:event_law}),
shows that 
$
T/(1-T\delta)>t_{k+1}-t_{k}>T/(1+T\delta).
$
By the selection of $\delta$ in (\ref{eq:delta}), one has 
\[
\delta^{*}>\frac{T^2\delta}{1-T\delta}>t_{k+1}-t_{k}-T>-\frac{T^2\delta}{1+T\delta}>-\delta^{*},
\]
which implies that $|t_{k+1}-t_{k}-T|<\delta^{*}$, $\forall k\geq k^{*}$.
Thus, the proof is complete. \eproof

\brem
For a linear function $\gamma$ and thus $
\bar \gamma$, the event-triggered law (\ref{eq:event_law}) directly leads to 
 a periodic sampling scheme.
 \erem

\subsection{Interconnected Systems\label{sub:IS}}

In this section, we consider event-triggered control of a nonlinear
interconnected system described  as follows 
\begin{align}
\dot{z}(t) & =  q(z(t),x(t),u(t)),\nonumber \\
\dot{x}(t) & =  f(z(t),x(t),u(t)),\label{eq:sys_zx}
\end{align}
where $z\in\mathbb{R}^{q}$ and $x\in\mathbb{R}^{n}$ are the states of the
two subsystems, and $u\in\mathbb{R}^{m}$ is the input. The {\blue continuous} functions $q:\mathbb{R}^{q}\times\mathbb{R}^{n}\times\mathbb{R}^{m}\to\mathbb{R}^{q}$
and $f:\mathbb{R}^{q}\times\mathbb{R}^{n}\times\mathbb{R}^{m}\to\mathbb{R}^{n}$
satisfy $q(0,0,0)=0$ and $f(0,0,0)=0$ such that $\mbox{col}(z,x)=0$
is the equilibrium point of the overall system with $u=0$. The state $z$ is
assumed not available for the feedback. {\color{blue}The aim  is to construct an event-triggered controller
(\ref{eq:sample_con}) such that 
%state stabilization with $\lim_{t\rightarrow\infty}\mbox{col}(z(t),x(t))=0$  
the system is globally
asymptotically stable at {\blue$\mbox{col}(z,x)=0$}
and Zeno-free behavior is achieved.} This problem was solved in \cite{Liu2015a}
using the cyclic small gain theorem. {\color{blue} Here, we will adopt the new event-triggered control scheme proposed
in Section \ref{sub:cd} to solve the problem and  explicitly characterize
how the $z$-dynamics affect the event-triggered control law. }

% provided that the controlled
%system has an ISS property from the sampling error %$e(t)=x(t_{k})-x(t)$
%to the state $x(t)$. 

%It also shows that the convergence rate of $z$-dynamics must be taken
%into account for the event-triggered law design even when only $x(t)$
%is used for the feedback. 

The closed-loop
system composed of (\ref{eq:sys_zx}) and (\ref{eq:sample_con}) can
be written as follows 
\begin{align}
\dot{z}(t) & =  q(z(t),x(t),g(x(t))-r(t)),\nonumber \\
\dot{x}(t) & =  f(z(t),x(t),g(x(t))-r(t)),\label{eq:closed-loop-zx}
\end{align}
with the auxiliary input $r(t)$ and output $\xi(t)$ defined as follows
\begin{equation}
r(t)  =  \int_{t_{k}}^{t}\xi(s)ds,\;t\in[t_{k},t_{k+1}),\;
\xi(t)  =  \frac{dg(x(t))}{dt}.\label{eq:r-2}
\end{equation}
The following ISS and bounded state and input to bounded output (BSIBO)
conditions are assumed for the closed-loop system (\ref{eq:closed-loop-zx}).
The assumptions could be matched through proper controller design in some real applications, 
for example, the specific design approach is discussed in Section~3 for lower-triangular systems.

\bass \label{ass:ass} The closed-loop system (\ref{eq:closed-loop-zx})
with {\color{blue}$r$} as the piecewise continuous bounded external input and {\color{blue}$\xi$}
as the output has following ISS properties.
\begin{itemize}
\item The $z$-dynamics and $x$-dynamics are ISS, i.e., 
\begin{align}
\|z(t)\|  \leq&  \max\{\beta_{z}(\|z(t_{0})\|,t-t_{0}),\gamma_{z}^{x}(\|x_{[t_{0},t]}\|),\nonumber\\
&\gamma_{z}^{r}(\|r_{[t_{0},t]}\|)\},\;\forall t>t_{0}\label{eq:ass_z}\\
\|x(t)\|  \leq&  \max\{\beta_{x}(\|x(t_{0})\|,t-t_{0}),\gamma_{x}^{z}(\|z_{[t_{0},t]}\|),\nonumber\\
&\gamma_{x}^{r}(\|r_{[t_{0},t]}\|)\},\;\forall t>t_{0}\label{eq:ass_x}
\end{align}
for some functions $\beta_{x},\beta_{z}\in\mathcal{KL}$ and $\gamma_{z}^{x},\gamma_{z}^{r},\gamma_{x}^{z},\gamma_{x}^{r}\in\mathcal{K}_{\infty}$.
\item It is BSIBO viewing $z$ and $x$ as states, $r$ as the input and
$\xi$ as the output, i.e.,
\begin{align}
\|\xi(t)\|\leq&\max\{\gamma_{\xi}^{z}(\|z_{[t_{0},t]}\|),\gamma_{\xi}^{x}(\|x_{[t_{0},t]}\|),\gamma_{\xi}^{r}(\|r_{[t_{0},t]}\|)\},\nonumber\\
&\qquad \qquad  \qquad \qquad \qquad \qquad  \forall t>t_{0}\label{eq:ass_xi}
\end{align}
for some functions $\gamma_{\xi}^{z},\gamma_{\xi}^{x},\gamma_{\xi}^{r}\in\mathcal{K}_{\infty}$.
\end{itemize}
\eass
A useful lemma is presented as follows.
\blem (\cite{Zhu2017}) \label{lem:zx} Consider the system (\ref{eq:closed-loop-zx})
with $\Lambda=[z\t,x\t]\t$. Suppose $r(t)$ is the piecewise continuous
bounded input. If  (\ref{eq:ass_z}) and (\ref{eq:ass_z}) hold
and the small gain condition $\gamma_{z}^{x}\circ\gamma_{x}^{z}(s)<s,\;\forall s>0$
is satisfied, the system (\ref{eq:closed-loop-zx}) is ISS in the
sense of
\begin{align}
\|z(t)\| & \leq  \max\{\bar{\beta}_{z}(\|\Lambda(t_{0})\|,t-t_{0}),\bar{\gamma}_{z}^{r}(\|r_{[t_{0},t]}\|)\},\label{eq:RISSZ}\\
\|x(t)\| & \leq  \max\{\bar{\beta}_{x}(\|\Lambda(t_{0})\|,t-t_{0}),\bar{\gamma}_{x}^{r}(\|r_{[t_{0},t]}\|)\},\label{eq:RISSX}
\end{align}
for all $t\geq t_0$
where  $\bar{\beta}_{\zeta},\bar{\beta}_{\chi}\in\mathcal{KL}$
and class $\mathcal{K}_{\infty}$ functions $\bar{\gamma}_{z}^{r}=\max\{\gamma_{z}^{r},\gamma_{z}^{x}\circ\gamma_{x}^{r}\},\;\bar{\gamma}_{x}^{r}=\max\{\gamma_{x}^{r},\gamma_{x}^{z}\circ\gamma_{z}^{r}\}.$
\elem 

\brem \label{rem:ISS} {\blue As} in \cite{Liu2015a}, we deliberately
consider the ISS property for the $z$ and $x$-dynamics separately rather
than consider that for $\Lambda=[z\t,x\t]\t$ as a whole.  On one hand, it facilitates examination of $z$
and $x$-dynamics' individual effect on the event-triggered control
design. On the other hand, under the small gain condition $\gamma_{z}^{x}\circ\gamma_{x}^{z}(s)<s,\;\forall s>0$,
we can derive the ISS property for  the $\Lambda$-dynamics 
\begin{equation}
\|\Lambda(t)\|\leq\max\{\beta_{\Lambda}(\|\Lambda(t_{0})\|,t-t_{0}),\gamma_{\Lambda}(\|r_{[t_{0},t]}\|)\},\label{eq:Lamda}
\end{equation}
for  all $t\geq t_0$
where   $\beta_{\Lambda}\in\mathcal{KL}$ and $\gamma_{\Lambda}:=2\max\{\bar{\gamma}_{z}^{r},\bar{\gamma}_{x}^{r}\}\\ \in\mathcal{K}_{\infty}$.
By Lemma \ref{lem:zx}, we can use the ISS properties (\ref{eq:RISSZ})
and (\ref{eq:RISSX}) with less conservative gain functions $\bar{\gamma}_{z}^{r}$
and $\bar{\gamma}_{x}^{r}$ instead of (\ref{eq:Lamda}) with $\gamma_{\Lambda}$
to design the event-triggered law. As will be explained in Remark\textcolor{red}{{}
}\ref{rem:sampling}, it may lead to a better sampling pattern.
\erem

\bthm \label{thm:etc-zx} Consider the system (\ref{eq:sys_zx})
with the controller (\ref{eq:sample_con}). Suppose Assumption \ref{ass:ass} is satisfied with the 
small gain condition $\gamma_{z}^{x}\circ\gamma_{x}^{z}(s)<s, \forall s>0.$
Let $\gamma:=\max\{\gamma_{\xi}^{z}\circ\bar{\gamma}_{z}^{r},\gamma_{\xi}^{x}\circ\bar{\gamma}_{x}^{r},\gamma_{\xi}^{r}\}$
where $\bar{\gamma}_{x}^{r}$ and $\bar{\gamma}_{z}^{r}$ are given
in Lemma~\ref{lem:zx}. Suppose $\gamma$ satisfies
\begin{equation}
\lim_{s\rightarrow0^{+}}\frac{\gamma(s)}{s}<\infty.\label{eq:glim-1}
\end{equation}
 Let $\epsilon>1$ and $\bar{\gamma}$ be a $\mathcal{K}_{\infty}$
function satisfying 
\begin{equation}
\bar{\gamma}(s)\geq\epsilon\gamma(s),\;\forall s>0\;\mbox{and}\;\lim_{s\rightarrow0^{+}}\frac{\bar{\gamma}(s)}{s}>0.\label{eq:gamma-2}
\end{equation}
The objectives of the event-triggered control  
are achieved  if the
event-triggered law (\ref{eq:event}) is 
\begin{gather}
t_{k+1}=\inf_{t>t_{k}}\{(t-t_{k})\bar{\gamma}(\|r_{[t_{k},t]}\|)>\|r_{[t_{k},t]}\|\},\;k\in \mathbb{N}\label{eq:event_law-1}
\end{gather}
where $r$ is define in (\ref{eq:r-2}).
\ethm

\proofnow By Lemma \ref{lem:zx}, one has (\ref{eq:RISSZ}) and (\ref{eq:RISSX}).
Following the similar argument in Theorem \ref{thm:etc}, we can prove
signals $x(t)$, $z(t)$, $r(t)$ and $\xi(t)$ of the closed-loop
system (\ref{eq:closed-loop-zx}) are bounded. Since all signals are
bounded, we can substitute (\ref{eq:RISSZ}) and (\ref{eq:RISSX})
into (\ref{eq:ass_xi}) and obtain 
\begin{equation}
\|\xi(t)\|\leq\max\{\beta(\|\Lambda(t_{0})\|,t-t_{0}),\gamma(\|r_{[t_{0},t]}\|)\}\label{eq:xi_b}
\end{equation}
for all $t\geq t_0$,
where $\beta\in\mathcal{KL}$. Note that (\ref{eq:RISSZ}), (\ref{eq:RISSX}) and
(\ref{eq:xi_b}) are similar to conditions of Theorem \ref{thm:etc}.
The rest of the proof directly follows that of Theorem \ref{thm:etc}.
\eproof

\brem Let us consider two special cases: (1) $x$ and $u$ do not
appear in the $z$-dynamics, i.e., $\dot{z}(t)=q(z(t)),\;\dot{x}(t)=f(x(t),z(t),u(t))$
($\gamma_{z}^{x}=\gamma_{z}^{r}=0$); (2) $u$ does not appear in
the $z$-dynamics but $x$ does, and $z$ does not appear in $x$-dynamics,
i.e, $\dot{z}(t)=q(z(t),x(t)),\;\dot{x}(t)=f(x(t),u(t))$ ($\gamma_{z}^{r}=\gamma_{x}^{z}=\gamma_{\xi}^{z}=0$).
It follows from Theorem \ref{thm:etc-zx} that $\bar{\gamma}(s)$
in (\ref{eq:gamma-2}) should be 
\[
\bar{\gamma}(s)\geq\epsilon\gamma(s):=\epsilon\max\{\gamma_{\xi}^{x}\circ\gamma_{x}^{r}(s),\gamma_{\xi}^{r}(s)\}.
\]
As opposed to the method in \cite{Liu2015a}, we explicitly show
that the variation of $z$-dynamics does not affect Zeno-free behavior
for both aforementioned cases. Specifically, in both cases, it is
not necessary to re-design the event-triggered law (\ref{eq:event_law-1})
when the $z$-dynamics vary. \erem

\brem \label{rem:sampling} For a function $\gamma$, it is
observed from (\ref{eq:event_law-1}) that less conservative selection
of $\bar{\gamma}(s)$ may increase the sampling interval $t_{k+1}-t_{k}$,
which could lead to a desirable sampling pattern that less number
of control executions are taken within a given period. If the ISS
property (\ref{eq:Lamda}) is used to derive an event-triggered law
rather than (\ref{eq:RISSZ}) and (\ref{eq:RISSX}), we can derive
the following inequality similar to (\ref{eq:xi_b}), 
\begin{equation}
\|\xi(t)\|\leq\max\{\beta(\|\Lambda(t_{0})\|,t-t_{0}),\tilde{\gamma}(\|r_{[t_{0},t]}\|)\},\label{eq:xi_b-2}
\end{equation}
for $t\geq t_0$
with 
$
\tilde{\gamma}:=\max\{\gamma_{\xi}\circ\gamma_{\Lambda},\gamma_{\xi}^{r}\}
$
where $\gamma_{\xi}=\max\{\gamma_{\xi}^{x},\gamma_{\xi}^{z}\}$ and
$\gamma_{\Lambda}:=2\max\{\bar{\gamma}_{z}^{r},\bar{\gamma}_{x}^{r}\}$.
It follows from the proof of Theorem \ref{thm:etc-zx} that $\bar{\gamma}(s)$
should be 
\begin{equation}
\bar{\gamma}(s)\geq\epsilon\tilde{\gamma}(s).\label{eq:gamma_s}
\end{equation}
The fact  $\tilde{\gamma}(s)>\gamma(s)$  makes the choice of $\bar{\gamma}(s)$
more conservative. \erem

% \brem 
% For linear systems, the output event-triggered  feedback controller is proposed based on Riccati equation  and LMIs \cite{borgers2018riccati,khashooei2017output,abdelrahim2018co}. For nonlinear systems,  the output event-triggered  control 
% is  formulated as the stabilization problem of a hybrid system and dissipativity like function is assumed, e.g., in \cite{dolk2017output,abdelrahim2017robust}.  The event-triggered law relies on the calculation of MASP $T$, which is designed to wait  $T$ units of time before it considers the event-triggering condition that utilizes both
% the actuation and the measurement error.  In this paper, we consider the event-triggered control via partial state feedback and  the controller design is based on Assumption \ref{ass:ass} with the ISS condition given  in  the max-form and the event-triggered law  only utilizes the actuation error. 
% \erem

\section{Lower-Triangular Systems \label{sec:lts}}
\subsection{Problem Formulation}
In this section, we consider the event-triggered control for a class of
lower-triangular systems
\begin{align}
\dot{z}_{j}(t) & =  q_{j}(\h{z}_{j}(t),\h{x}_{j}(t),w),\nonumber \\
\dot{x}_{j}(t) & =  f_{j}(\h{z}_{j}(t),\h{x}_{j}(t),w)+b_{j}x_{j+1},\;j=1,\cdots,\ell\label{eq:LTS}
\end{align}
where $\h{z}_{j}:=\mbox{col}(z_{1},\cdots,z_{j})$ 
and $\h{x}_{j}:=\mbox{col}(x_{1},\cdots,x_{j})$
are the states, $u:=x_{\ell+1}$ is the control input, $b_{j}$'s
are constants and $\ell$ is the relative degree. Note that  $z_{j}\in\mathbb{R}^{n_{j}}$ and   $x_{j}\in\mathbb{R}$.  $w\in\mathbb{W}$
represents system uncertainties (such as unknown parameters) in a
known compact set $\mathbb{W}$, while $\h{z}_{j}$-dynamics are called
dynamic uncertainties where the function $q_{j}$'s are not precisely
known and $\h{z}_{j}$ is not available for feedback. Note that  $w$ can be time-varying. The functions $q_{j}$
and $f_{j}$ are assumed to be sufficiently smooth and satisfy $q_{j}(0,0,w)=0$
and $f_{j}(0,0,w)=0$. The continuous-time
stabilization of such system has been solved using the Lyapunov function
method \cite{Chen2004} and the small gain theorem \cite{Jiang1997} based
on backstepping technique. In the spirit of backstepping, we introduce
the coordinate transformation 
\begin{align}
\bar{x}_{1} & =  x_{1}\nonumber \\
\bar{x}_{j} & =  x_{j}-\vartheta_{j-1}(\bar{x}_{j-1}),\;j=2,\cdots,\ell+1\label{eq:CT}
\end{align}
where the functions $\vartheta_{1},\cdots,\vartheta_{\ell}$ satisfying
$\vartheta_{i}(0)=0$ are virtual controllers to be designed at each
recursive step. Under the coordinate (\ref{eq:CT}), the system (\ref{eq:LTS})
becomes 
\begin{align}
\dot{z}_{j}(t) & =  \bar{q}_{j}(\h{z}_{j}(t),\h{\bar{x}}_{j}(t),w)\nonumber \\
\dot{\bar{x}}_{j}(t) & =  \bar{f}_{j}(\h{z}_{j}(t),\h{\bar{x}}_{j}(t),w)+\vartheta_{j}(\bar{x}_{j}(t))+\bar{x}_{j+1}(t),\nonumber\\
&\quad \quad \quad \quad  j=1,\cdots,\ell\label{eq:zxbar}
\end{align}
where $\h{\bar{x}}_{j}:=\mbox{col}(\bar{x}_{1},\cdots,\bar{x}_{j})$,
$\bar{q}_{j}(\h{z}_{j},\h{\bar{x}}_{j},w)$ and $\bar{f}_{j}(\h{z}_{j},\h{\bar{x}}_{j},w)$
are sufficiently smooth functions satisfying $\bar{q}_{j}(0,0,w)=0$
and $\bar{f}_{j}(0,0,w)=0$. For continuous-time stabilization,
the controller is of the form {\blue$u=x_{\ell+1}=\vartheta_{\ell}(\bar{x}_{\ell})$}
by setting {\blue$\bar{x}_{\ell+1}=0$}. For event-triggered control,
we adopt the method developed in Section \ref{sec:etc} and
propose the controller as follows 
\begin{equation}
u(t)=\vartheta_{\ell}(\bar{x}_{\ell}(t_{k}))\mbox{ or }u(t)=\vartheta_{\ell}(\bar{x}_{\ell}(t))-\bar{x}_{\ell+1}(t)\label{eq:control}
\end{equation}
where $\bar{x}_{\ell+1}(t)$ is the sampling error defined as 
\begin{equation}
\bar{x}_{\ell+1}(t)=\int_{t_{k}}^{t}\xi(s)ds,\;t\in[t_{k},t_{k+1}),\;\xi(t)=\frac{d\vartheta_{\ell}(\bar{x}_{\ell}(t))}{dt}.\label{eq:r-1}
\end{equation}
with the triggering law to be designed as
\begin{equation}
t_{k+1}=\inf_{t>t_k}\{\Xi(t_k,\h{\bar{x}}_{\ell}(t_{k}),\h{\bar{x}}_{\ell}(t))>0\},\;k\in \mathbb{N}.\label{eq:event-1}
\end{equation}
The backstepping controller design (\ref{eq:CT}) is a classic  continuous-time stabilization  technique for lower-triangular nonlinear systems. For event-triggered control, the continuous-time controller $u(t)=\vartheta_{\ell}(\bar{x}_{\ell}(t))$ needs to be designed such that the closed-loop system has an ISS gain from state sampling error $e(t):=\h{x}_{\ell}(t)-\h{x}_{\ell}(t_k)$  to states $\h{x}_{\ell}$  and $\h{z}_{\ell}$  for existing event-triggered control design or from actuation error  $\bar{x}_{\ell+1}(t)$ to states for the  controller in this paper. We will design a new recursive recipe to separate the ISS properties of $\h{\bar{x}}_{\ell}$ and  $\h{z}_{\ell}$ for the closed-loop system  (\ref{eq:zxbar}) and employ Theorem \ref{thm:etc-zx} to design the event-triggered
law (\ref{eq:event-1}). The traditional treatment in the literature
mixes $\h{z}_{j}$ and $\h{\bar{x}}_{j}$-dynamics and derives the
ISS property for $\|\mbox{col}(\h{z}_{j}(t),\h{\bar{x}}_{j}(t))\|$
as a whole at each recursive step. In the spirit of the design method in Section
\ref{sub:IS}, we separately consider the $\h{z}_{j}$ and $\h{\bar{x}}_{j}$-dynamics
at each step to derive the following ISS properties. In particular, we aim to 
design the continuous-time controllers $\vartheta_{1},\cdots,\vartheta_{\ell}$
such that the following statement holds
for $j=1,\cdots,\ell$.

\textit{Statement} $j$: The $(\h{z}_{j},\h{\bar{x}}_{j})$-dynamics of (\ref{eq:zxbar}) are
ISS in the sense of 
\begin{align}
\|\h{z}_{j}(t)\| & \leq  \max\{\bar{\beta}_{Z_{j}}(\|\mbox{col}(\h{z}_{j}(t_{0}),\h{\bar{x}}_{j}(t_{0}))\|,t-t_{o}),\nonumber\\
&\quad \bar{\gamma}_{Z_{j}}^{x_{j+1}}(\|\bar{x}_{j+1}\|)\},\nonumber \\
\|\h{\bar{x}}_{j}(t)\| & \leq  \max\{\bar{\beta}_{X_{j}}(\|\mbox{col}(\h{z}_{j}(t_{0}),\h{\bar{x}}_{j}(t_{0}))\|,t-t_{o}),\nonumber\\
&\quad \bar{\gamma}_{X_{j}}^{x_{j+1}}(\|\bar{x}_{j+1}\|)\},\;\forall t\geq t_{0}\label{eq:cond}
\end{align}
for some functions $\bar{\beta}_{Z_{j}},\bar{\beta}_{X_{j}}\in\mathcal{KL}$
and $\bar{\gamma}_{Z_{j}}^{x_{j+1}},\bar{\gamma}_{X_{j}}^{x_{j+1}}\in\mathcal{K}_\infty$
to be calculated. Furthermore, the gain functions satisfy 
\begin{equation}
\bar{\gamma}_{X_{j}}^{x_{j+1}}(s)<s\;\mbox{ and }\lim_{s\rightarrow0^{+}}\frac{\bar{\gamma}_{Z_{j}}^{x_{j+1}}(s)}{s}<\infty,\;\forall s>0.\label{eq:lim_con}
\end{equation}

\subsection{Recursive Controller Design}\label{sub:rec}

We will propose the event-triggered controller under a standard assumption.

\bass \label{ass:zj} The $z_{j}$-subsystem for $j=1,\cdots,\ell$
in (\ref{eq:LTS}) is ISS  for all $w\in\mathbb{W}$ viewing $z_{j}$
as state, $\h{z}_{j-1}$ and $\h{x}_{j}$ as inputs with the ISS functions
$\gamma_{z_{j}}^{z},\gamma_{z_{j}}^{x}\in\mathcal{K}_{\infty}$ satisfying $\lim_{s\rightarrow0^{+}}\gamma_{z_{j}}^{z}(s)/s<\infty$
 and   $\lim_{s\rightarrow0^{+}}\gamma_{z_{j}}^{x}(s)/s<\infty$,
respectively. \eass

\brem Under Assumption \ref{ass:zj}, the $z_{j}$-subsystem
in (\ref{eq:zxbar}) is ISS viewing $z_{j}$ as state and $\mbox{col}(\h{z}_{j-1},\h{\bar{x}}_{j})$
as input, in particular, 
\begin{align}
\|z_{j}(t)\|\leq&\max\big\{\beta_{z_{j}}(|z_{j}(t_{0})|,t-t_{0}),\gamma_{z_{j}}^{z}(\|\h{z}{}_{j-1[t_{0},t]}\|),\nonumber\\
&\quad \bar{\gamma}_{z_{j}}^{x}(\|\h{\bar{x}}_{j[t_{0},t]}\|)\big\},\;\forall t>t_{0}\label{eq:zi_iss-1}
\end{align}
for some functions $\beta_{z_{j}}\in\mathcal{KL}$ and 
$\bar{\gamma}_{z_{j}}^{x} \in\mathcal{K}_{\infty}$ with $\lim_{s\rightarrow0^{+}}\bar{\gamma}_{z_{j}}^{x}(s)/s<\infty.$
\erem

For $w\in\mathbb{W}$, direct calculation leads to 
\begin{equation}
|\bar{f}_{1}(z_{1},\bar{x}_{1},w)|\leq\iota_{1}(|z_{1}|)|z_{1}|+m_{1}(\bar{x}_{1})|\bar{x}_{1}|,\label{eq:f1}
\end{equation}
and, for $j=2,\cdots,\ell$,
\begin{align}
|\bar{f}_{j}(\h{z}_{j},\h{\bar{x}}_{j},w)|<&\iota_{j}(\|\h{z}_{j}\|)\|\h{z}_{j}\|+\tilde{m}_{j}(\|\h{\bar{x}}_{j-1}\|)\|\h{\bar{x}}_{j-1}\|\nonumber\\
&+m_{j}(\bar{x}_{j})|\bar{x}_{j}|,\label{eq:fj}
\end{align}
where we select non-negative sufficiently smooth functions $\iota_{j}$,
$m_{j}$ and $\tilde{m}_{j}$. Denote 
\begin{gather*}
\bar{\iota}_{j}(s)=\frac{1}{4c_{j}}\iota_{j}^{2}(s)s^{2},\;j=1,\cdots,\ell\\
\bar{c}_{1}=c_{1},\;\bar{c}_{j}=2c_{j},\;\bar{m}_{j}(s)=\frac{1}{4c_{j}}\tilde{m}_{j}^{2}(s)s^{2},\;j=2,\cdots,\ell
\end{gather*}
for constants $c_{j}>0$. Define 
$
\gamma_{Z_{j}}^{X_{j}}=2\max\{\bar{\gamma}_{z_{j}}^{x},\bar{\gamma}_{Z_{j-1}}^{x_{j}},\gamma_{z_{j}}^{z}\circ\bar{\gamma}_{Z_{j-1}}^{x_{j}}\},\;j=2,\cdots,\ell.
$
The continuous-time stabilization controller is designed as follows
\begin{equation}
\vartheta_{j}(\bar{x}_{j})=-(\bar{c}_{j}+\frac{1}{4}b_{j}^{2}+m_{j}(\bar{x}_{j})+\psi_{j}(|\bar{x}_{j}|))\bar{x}_{j}/b_{j},\;j=1,\cdots,\ell\label{eq:nu}
\end{equation}
where we select some sufficiently smooth and even functions $\psi_{j}(s)$
satisfying the following conditions 
\begin{gather}
\psi_{1}(s)>k_{1}\max\{\bar{\iota}_{1}\circ\gamma_{z_{1}}^{x}(s)/s^{2},1\}\nonumber \\
\psi_{j}(s)>k_{j}\max\{\bar{m}_{j}(s)/s^{2},\bar{\iota}_{j}\circ\gamma_{Z_{j}}^{X_{j}}(2s)/s^{2},2\}\label{eq:psi}
\end{gather}
for $k_{1}>2$ and $k_{j}>3,\;j=2,\cdots,\ell$.

\blem \label{lem:ZX}Consider the closed-loop system composed of
(\ref{eq:LTS}) and (\ref{eq:control}) with any piecewise bounded
external input $\bar{x}_{j+1}$. Let $\vartheta_{j}(\bar{x}_{j})$, 
for $j=1,\cdots,\ell$, be given in (\ref{eq:nu}). Then, Statement
$j$ holds for $j=1,\cdots,\ell$. \elem

\proofnow   We will use mathematical induction to prove
Statement $j$. For step $j=1$. Let $V_{1}(\bar{x}_{1})=\bar{x}_{1}^{2}/2$
be a Lyapunov function candidate for $\bar{x}_{1}$-subsystem. Then,
one has 
\begin{align*}
\dot{V}_{1}(\bar{x}_{1}) & =  \bar{x}_{1}[f_{1}(z_{1},\bar{x}_{1},w)+\bar{x}_{2}+\vartheta_{1}(\bar{x}_{1})]\\
 & \leq  -\rho_{1}(|\bar{x}_{1}|)+\bar{\iota}_{1}(|z_{1}|)+\bar{x}_{2}^{2}
\end{align*}
for a class $\mathcal{K}_\infty$ function $\rho_{1}(s)=\psi_{1}(s)s^{2}$
where we used (\ref{eq:nu}) and (\ref{eq:f1}). As a result, the
$\bar{x}_{1}$-dynamics are ISS viewing $\bar{x}_{1}$ as state and
$\mbox{col}(z_{1},\bar{x}_{2})$ as input, in particular, 
\begin{align}
|\bar{x}_{1}(t)|\leq&\max\big\{\beta_{x_{1}}(|\bar{x}_{1}(t_{0})|,t-t_{0}),\gamma_{x_{1}}^{z}(\|z_{1}\|),\nonumber\\
&\gamma_{x_{1}}^{x_{2}}(\|\bar{x}_{2}\|)\big\},\forall t>t_{0}\label{eq:x1_ineq}
\end{align}
where $\gamma_{x_{1}}^{z}=\rho_{1}^{-1}\circ(k_{1}\bar{\iota}_{1})$
and $\gamma_{x_{1}}^{x_{2}}(s)=\rho_{1}^{-1}(k_{1}s^{2})$. According
to (\ref{eq:zi_iss-1}), we have 
\begin{equation}
\|z_{1}(t)\|\leq\max\left\{ \beta_{z_{1}}(\|z_{1}(t_{o})\|,t-t_{0}),\bar{\gamma}_{z_{1}}^{x}(\|\bar{x}_{1}\|)\right\} ,\forall t>t_{0}.\label{eq:z1_ineq}
\end{equation}
Choosing $\psi_{1}(s)$ in (\ref{eq:psi}) leads to the small gain
condition 
$
\gamma_{x_{1}}^{z}\circ\gamma_{z_{1}}^{x}(s)=\rho_{1}^{-1}\circ(k_{1}\bar{\iota}_{1})\circ\bar{\gamma}_{z_{1}}^{x}(s)<s
$
and $\gamma_{x_{1}}^{x_{2}}(s)=\rho_{1}^{-1}(k_{1}s^{2})<s.$ By Lemma
\ref{lem:zx}, the $(z_{1},\bar{x}_{1})$-dynamics are ISS in the
sense of (\ref{eq:cond}) with 
$
\bar{\gamma}_{X_{1}}^{x_{2}}=\gamma_{x_{1}}^{x_{2}},\;\bar{\gamma}_{Z_{1}}^{x_{2}}=\bar{\gamma}_{z_{1}}^{x}\circ\gamma_{x_{1}}^{x_{2}}.
$
Due to $\lim_{s\rightarrow0^{+}}\bar{\gamma}_{z_{1}}^{x}(s)/s<\infty,$
one can find the function $\rho_{1}\in\mathcal{K}$ and the function
$\psi_{1}(s)$ satisfying (\ref{eq:psi}). Also, one has $\lim_{s\rightarrow0^{+}}\bar{\gamma}_{Z_{1}}^{x_{2}}(s)/s<\infty$,
i.e., (\ref{eq:lim_con}) is satisfied for $j=1$. Statement $j$
holds for $j=1$.

For step $j\geq2$. It is noted that the $(\h{z}_{j},\h{\bar{x}}_{j})$-system
is composed of the $(\h{z}_{j-1},\h{\bar{x}}_{j-1})$-subsystem, the
$\bar{x}_{j}$-subsystem, and the $z_{j}$-subsystem. For the purpose
of induction, suppose that $\vartheta_{j-1}(\bar{x}_{j-1})$ has been
designed such that Statement $j-1$ holds. Then, we aim to design
$\vartheta_{j}(\bar{x}_{j})$ in this step such that Statement $j$
also holds. Let $V_{j}=\bar{x}_{j}^{2}/2$ be a Lyapunov function
candidate for $\bar{x}_{j}$-subsystem. Then, one has 
\begin{gather*}
\dot{V}_{j}(\bar{x}_{j})=\bar{x}_{j}[\bar{f}_{j}(\h{z}_{j},\h{\bar{x}}_{j},w)+\bar{x}_{j+1}+\nu_{j}(\bar{x}_{j})]\\
\leq-\rho_{j}(\bar{x}_{j})+\bar{\iota}_{j}(\|\h{z}_{j}\|)+\bar{m}_{j}(\|\h{\bar{x}}_{j-1}\|)+\bar{x}_{j+1}^{2}
\end{gather*}
for a class $\mathcal{K}_\infty$ function $\rho_{j}(s)=\psi_{j}(s)s^{2}$
where we used (\ref{eq:nu}) and (\ref{eq:fj}). As a result, the
$\bar{x}_{j}$-dynamics are ISS viewing $\bar{x}_{j}$ as state and
$\mbox{col}(\h{z}_{j},\h{\bar{x}}_{j-1},\bar{x}_{j+1})$ as input,
in particular, 
\begin{align}
|\bar{x}_{j}(t)|\leq&\max\big\{\beta_{x_{j}}(|x_{j}(t_{0})|,t-t_{0}),\gamma_{x_{j}}^{z}(\|\h{z}_{j}\|),\gamma_{x_{j}}^{x}(\|\h{\bar{x}}_{j-1}\|),\nonumber\\
&\gamma_{x_{j}}^{x_{j+1}}(|\bar{x}_{j+1}|)\big\},\;\forall t>t_{0}\label{eq:xj_ineq}
\end{align}
where $\gamma_{x_{j}}^{z}=\rho_{j}^{-1}\circ(k_{j}\bar{\iota}_{j}),\;\gamma_{x_{j}}^{x}=\rho_{j}^{-1}\circ(k_{j}\bar{m}_{j}),\;\gamma_{x_{j}}^{x_{j+1}}(s)=\rho_{j}^{-1}(k_{j}s^{2}).$
According to (\ref{eq:zi_iss-1}), one has 
\begin{align}
\|z_{j}(t)\|\leq&\max\big\{\bar{\beta}_{z_{j}}(|z_{j}(t_{0})|,t-t_{0}),\gamma_{z_{j}}^{z}(\|\h{z}_{j-1}\|),\bar{\gamma}_{z_{j}}^{x}(\|\h{\bar{x}}_{j}\|)\big\},\nonumber\\
&\qquad \qquad \qquad \qquad \qquad \qquad  \forall t>t_{0}.\label{eq:zj_ineq}
\end{align}
Consider class $\mathcal{K}$ functions $\bar{\gamma}_{Z_{j-1}}^{x_{j}}$,
$\bar{\gamma}_{Z_{j-1}}^{\sigma}$, $\bar{\gamma}_{X_{j-1}}^{x_{j}}$,
and $\bar{\gamma}_{X_{j-1}}^{\sigma}$ from Statement $j-1$. Choosing
$\psi_{j}(s)$ in (\ref{eq:psi}) leads to small gain condition 
$
\gamma_{x_{j}}^{x}\circ\bar{\gamma}_{X_{j-1}}^{x_{j}}(s)=\rho_{j}^{-1}\circ(k_{j}\bar{m}_{j})\circ\bar{\gamma}_{X_{j-1}}^{x_{j}}(s)\leq\rho_{j}^{-1}\circ(k_{j}\bar{m}_{j})(s)<s
$
by noting $\bar{\gamma}_{X_{j-1}}^{x_{j}}(s)<s$. Define 
\begin{align*}
\gamma_{Z_{j}}^{X_{j}} & =  2\max\{\bar{\gamma}_{z_{j}}^{x},\bar{\gamma}_{Z_{j-1}}^{x_{j}},\gamma_{z_{j}}^{z}\circ\bar{\gamma}_{Z_{j-1}}^{x_{j}}\}\\
\gamma_{X_{j}}^{Z_{j}} & =  2\gamma_{x_{j}}^{z},\;\gamma_{X_{j}}^{x_{j+1}}=2\gamma_{x_{j}}^{x_{j+1}}.
\end{align*}
Using (\ref{eq:psi}), one has the other small gain condition 
\[
\gamma_{Z_{j}}^{X_{j}}\circ\gamma_{X_{j}}^{Z_{j}}(s)=\gamma_{Z_{j}}^{X_{j}}\circ(2\gamma_{x_{j}}^{z})(s)=\gamma_{Z_{j}}^{X_{j}}\circ(2\rho_{j}^{-1})\circ(k_{j}\bar{\iota}_{j})(s)<s
\]
By Lemma 3.2 in \cite{Zhu2017}, the $(\h{z}_{j},\h{\bar{x}}_{j})$-dynamics
are ISS in the sense of (\ref{eq:cond}), with 
\[
\bar{\gamma}_{Z_{j}}^{x_{j+1}}=\gamma_{Z_{j}}^{X_{j}}\circ\gamma_{X_{j}}^{x_{j+1}},\;\bar{\gamma}_{X_{j}}^{x_{j+1}}=\gamma_{X_{j}}^{x_{j+1}},
\]
and $\bar{\gamma}_{X_{j}}^{x_{j+1}}(s)=2\rho_{j}^{-1}(k_{j}s^{2})<s$
due to (\ref{eq:psi}). Due to $\lim_{s\rightarrow0^{+}}\bar{\gamma}_{z_{j}}^{x}(s)/s<\infty,$
$\lim_{s\rightarrow0^{+}}\bar{\gamma}_{Z_{j-1}}^{x_{j}}(s)/s<\infty,$
and $\lim_{s\rightarrow0^{+}}\gamma_{z_{j}}^{z}(s)/s<\infty,$ one
has $\lim_{s\rightarrow0^{+}}\gamma_{Z_{j}}^{X_{j}}(s)/s<\infty.$
One can always find a function $\rho_{i}\in\mathcal{K}_\infty$ and a function
$\psi_{j}(s)$ satisfying (\ref{eq:psi}). Also, one can verify (\ref{eq:lim_con})
for Statement $j$. The proof for Statement $j=2,\cdots,\ell$ is
thus complete. \eproof
\subsection{Event-Triggered Control}
Note that Statement $\ell$ implies that 
\begin{align}
\|\h{z}_{\ell}(t)\|  \leq&  \max\{\bar{\beta}_{Z_{\ell}}(\|\mbox{col}(\h{z}_{\ell}(t_{0}),\h{\bar{x}}_{\ell}(t_{0}))\|,t-t_{o}),\nonumber\\
& \bar{\gamma}_{Z_{\ell}}^{x_{\ell+1}}(\|\bar{x}_{\ell+1}\|)\},\nonumber \\
\|\h{\bar{x}}_{\ell}(t)\|  \leq & \max\{\bar{\beta}_{X_{\ell}}(\|\mbox{col}(\h{z}_{\ell}(t_{0}),\h{\bar{x}}_{\ell}(t_{0}))\|,t-t_{o}),\nonumber\\
&\bar{\gamma}_{X_{\ell}}^{x_{j+1}}(\|\bar{x}_{\ell+1}\|)\},\;\forall t\geq t_{0} . \label{eq:cond-1}
\end{align}
\brem
Note from (\ref{eq:cond-1}) that $\bar{x}_{\ell+1}=\vartheta_{\ell}(\bar{x}_{\ell}(t))-\vartheta_{\ell}(\bar{x}_{\ell}(t_k))=\bar\vartheta_{\ell}(\h{x}_{\ell}(t_k)+e)-\bar\vartheta_{\ell}(\h{x}_{\ell}(t_k))$  for some function $\bar\vartheta$ where $e(t):=\h{x}_{\ell}(t)-\h{x}_{\ell}(t_k)$. Then, there exists a class $\mathcal{K}$ function $\tilde \vartheta$ such that $\bar{x}_{\ell+1}\leq \tilde \vartheta(\|q\|) $. Therefore, the recursive controller design presented in Section \ref{sub:rec} can also render the closed-loop system have an ISS gain from state sampling error $e$ to  $\h{x}_{\ell}$  and $\h{z}_{\ell}$. By separating ISS gains from    $e$ to  $\h{x}_{\ell}$  and $\h{z}_{\ell}$, it is possible to design the event-triggered controller that only utilized partial state  $\h{x}_{\ell}$, as was done in \cite{Liu2015a}.
\erem
Therefore, the ISS property from the auxiliary input $\bar{x}_{\ell+1}$
to $\mbox{col}(\h{z}_{\ell}(t),\h{\bar{x}}_{\ell}(t))$ is achieved,
which is similar to have the conditions (\ref{eq:ass_z}) and (\ref{eq:ass_x})
in Assumption \ref{ass:ass} satisfied, by applying Lemma \ref{lem:zx}. We
also need to verify (\ref{eq:ass_xi}) in order to use Theorem \ref{thm:etc-zx}.
For this purpose, let us examine the auxiliary output $\xi$ in (\ref{eq:r-1}).
Note that 
\[
\xi(t)=\frac{d\vartheta_{\ell}(\bar{x}_{\ell})}{d\bar{x}_{\ell}}\dot{\bar{x}}_{\ell}=\alpha(\bar{x}_{\ell})[\bar{f}_{\ell}(\h{z}_{\ell},\h{\bar{x}}_{\ell},w)+\bar{x}_{\ell+1}+\vartheta_{\ell}(\bar{x}_{\ell})]
\]
where $\alpha(\bar{x}_{\ell}):=\frac{d\vartheta_{\ell}(\bar{x}_{\ell})}{d\bar{x}_{\ell}}$.
From $\vartheta_{\ell}(\bar{x}_{\ell})$ in (\ref{eq:nu}), one has
\begin{align*}
\alpha(\bar{x}_{\ell})  =&  -(\bar{c}_{\ell}+\frac{1}{4}b_{\ell}^{2}+m_{\ell}(\bar{x}_{\ell})+\psi_{\ell}(|\bar{x}_{\ell}|))/b_{\ell}\nonumber\\
&+\left(\frac{dm_{\ell}(\bar{x}_{\ell})}{d\bar{x}_{\ell}}+\frac{d\psi_{\ell}(|\bar{x}_{\ell}|)}{d\bar{x}_{\ell}}\right)\bar{x}_{\ell}/b_{\ell}
 =  c+\kappa_{1}(\bar{x}_{\ell})
\end{align*}
for some constant $c$ and sufficiently smooth function $\kappa_{1}$
satisfying $\kappa_{1}(0)=0$. Since $\bar{f}_{\ell}(\h{z}_{\ell},\h{\bar{x}}_{\ell},w)$
is sufficiently smooth, one has
\begin{align}
&|\bar{f}_{\ell}(\h{z}_{\ell},\h{\bar{x}}_{\ell},w)+\vartheta_{\ell}(\bar{x}_{\ell})+\bar{x}_{\ell+1}|<\nonumber\\
&\kappa_{2}(\|\h{z}_{\ell}\|)\|\h{z}_{\ell}\|+\kappa_{3}(\|\h{\bar{x}}_{\ell}\|)\|\h{\bar{x}}_{\ell}\|+|\bar{x}_{\ell+1}|\label{eq:fj-1}
\end{align}
for some non-negative sufficiently smooth functions $\kappa_{2}$
and $\kappa_{3}$. As a result, there exist $\gamma_{\xi}^{z},\gamma_{\xi}^{x}\in\mathcal{K}_{\infty}$
satisfying $\lim_{s\rightarrow0^{+}}\gamma_{\xi}^{z}(s)/s<\infty$
and $\lim_{s\rightarrow0^{+}}\gamma_{\xi}^{x}(s)/s<\infty$ such that
$
\|\xi(t)\|\leq\max\{\gamma_{\xi}^{z}(\|\h{z}_{\ell}\|),\gamma_{\xi}^{x}(\|\h{\bar{x}}_{\ell}\|),\gamma_{\xi}^{r}(|\bar{x}_{\ell+1}|)\},
$
where $\gamma_{\xi}^{r}(|\bar{x}_{\ell+1}|)=\alpha|\bar{x}_{\ell+1}|^{2}+\beta|\bar{x}_{\ell+1}|$
for some constants $\alpha,\beta>0$. It thus verifies (\ref{eq:ass_xi}).
Finally, we can check that 
$\lim_{s\rightarrow0^{+}}\gamma_{\xi}^{z}(s)/s<\infty$, $\lim_{s\rightarrow0^{+}}\gamma_{\xi}^{x}(s)/s<\infty$, $
\lim_{s\rightarrow0^{+}}\bar{\gamma}_{Z_{j}}^{x_{j+1}}(s)/s<\infty$, $\lim_{s\rightarrow0^{+}}\bar{\gamma}_{X_{j}}^{x_{j+1}}(s)/s<\infty
$
and hence $  \lim_{s\rightarrow0^{+}}\gamma(s)/s<\infty. $

By Theorem \ref{thm:etc-zx}, the conclusion on the event-triggered controller
is drawn as follows.

\bthm \label{thm:lts} Consider the closed-loop system composed of
(\ref{eq:LTS}) and (\ref{eq:control}) where $\vartheta_{j}(\bar{x}_{j})$,
for $j=1,\cdots,\ell$, is given in (\ref{eq:nu}).  Suppose  the closed-loop system satisfies Assumption \ref{ass:zj}. Let $\epsilon>1$
and $\bar{\gamma}\in\mathcal{K}_{\infty}$ be 
\begin{align}
\bar{\gamma}(s)\geq\epsilon\gamma(s),\;\forall s>0\; \label{eq:gamma-1}
\end{align}
where
$
\gamma(s):=\max\{\gamma_{\xi}^{z}\circ\bar{\gamma}_{Z_{j}}^{x_{j+1}}(s),
\gamma_{\xi}^{x}\circ\bar{\gamma}_{X_{j}}^{x_{j+1}}(s),\gamma_{\xi}^{r}(s)\}
$
and $\bar{\gamma}_{Z_{j}}^{x_{j+1}},\bar{\gamma}_{X_{j}}^{x_{j+1}}$
are given in (\ref{eq:cond}).The objectives of event-trigger control are achieved if the event-triggered law in (\ref{eq:event-1})
is 
\begin{align}
t_{k+1}=\inf_{t>t_{k}}\{(t-t_{k})\bar{\gamma}(\|r_{[t_{k},t]}\|)> \|r_{[t_{k},t]}\|\}, \;k\in \mathbb{N}.\label{eq:t_bar-1}
\end{align}
 \ethm

\section{Numerical Simulation \label{sec:ns}}

\begin{figure}[t]
\vspace{-1.6cm}
\centering\includegraphics[scale=1.0]{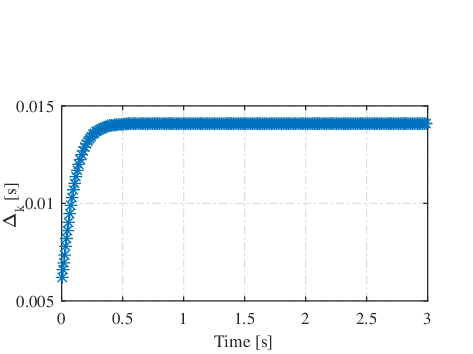}
\caption{Triggering instants under the designed scheme for
the system (\ref{eq:xzxz}).\label{fig:sim2}}
\end{figure}

\begin{figure}
\centering\includegraphics[scale=0.4]{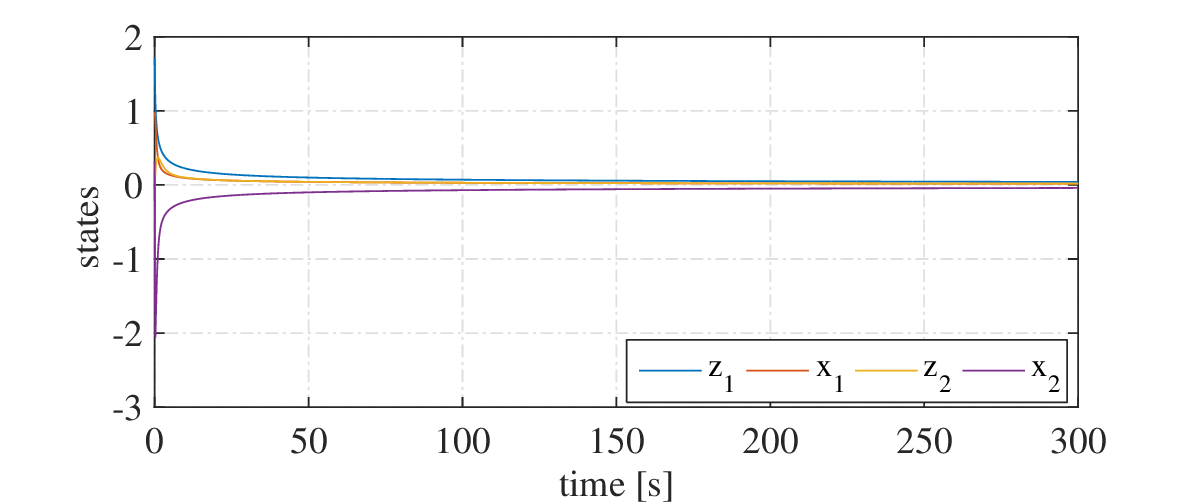}
\caption{Simulation profiles of the event-triggered control for
the system (\ref{eq:xzxz})\label{fig:sim1}}
\end{figure}

\begin{figure}
\centering\includegraphics[scale=0.38]{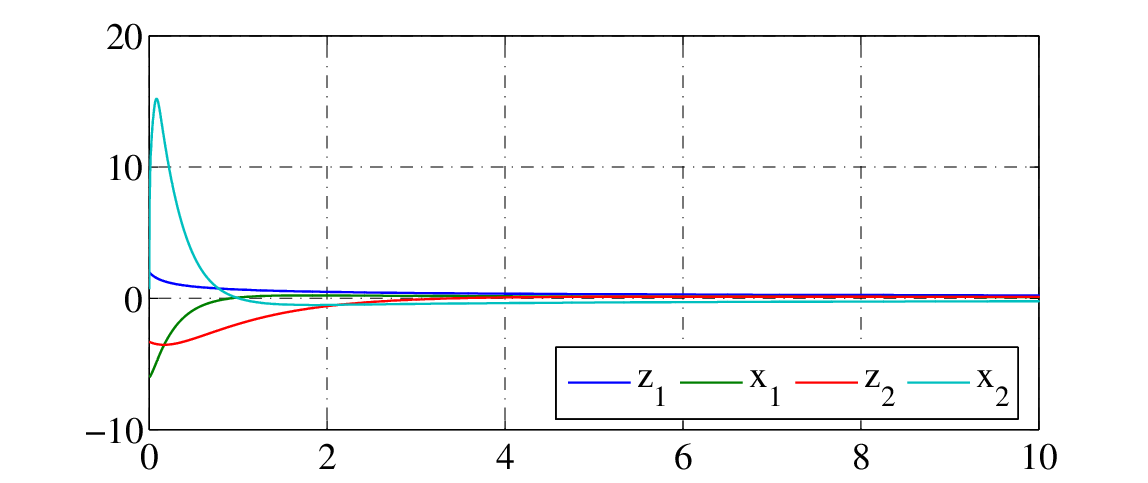}
\centering\includegraphics[scale=0.38]{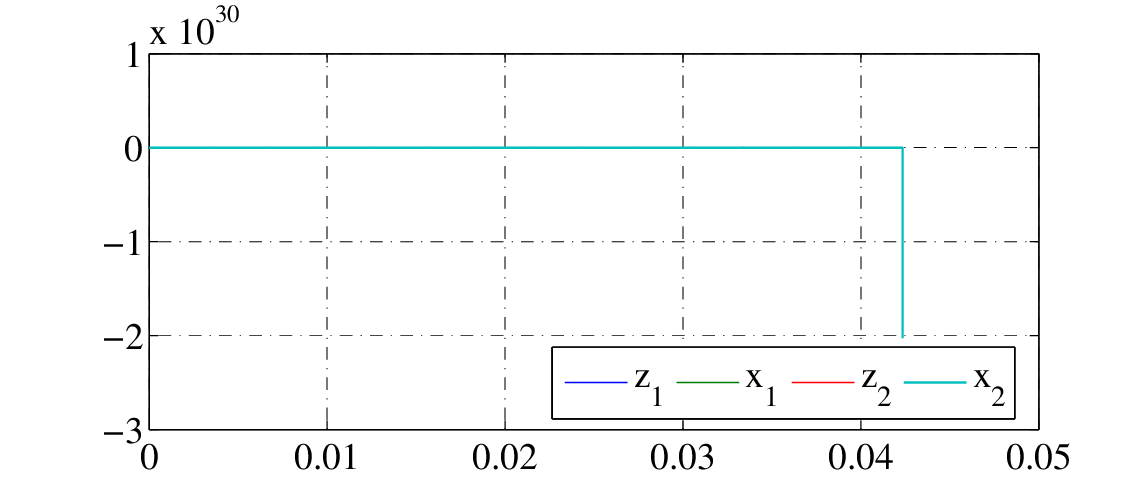}
\caption{The simulation comparison  in Section \ref{sec:ns}: proposed event-triggered law (upper) and periodic sampling with sampling interval being 0.0141 (lower) for  initial condition  $[z_1,x_1,z_2,x_2]\t=[-19.67,23.09,-24.77,13.755]\t$.\label{fig:sim3}}
\end{figure}

Consider the following lower-triangular system 
\begin{align}
\dot{z}_{1}(t) & =  -z_{1}^{3}(t)\nonumber \\
\dot{x}_{1}(t) & =  w_{1}z_{1}(t)+x_{2}(t)\nonumber \\
\dot{z}_{2}(t) & =  -z_{2}(t)+x_{1}(t)\nonumber \\
\dot{x}_{2}(t) & =  -w_{2}x_{1}(t)x_{2}(t)+x_{1}(t)+3u\label{eq:xzxz}
\end{align}
where $[z_{1},x_{1},z_{2},x_{2}]\t$ are states and $w_{1},w_{2}\in[0,1]$
are the parameter uncertainties. The $z_{1}$ and $z_{2}$-dynamics represent
dynamic uncertainties. We will use Theorem \ref{thm:lts} to design
the event-triggered controller. Applying Lemma \ref{lem:ZX}, the
continuous-time stabilization controller is designed as {\blue
$u=\vartheta_{2}(\bar{x}_{2})=-(0.3\bar{x}_{2}^{2}+5)\bar{x}_{2}
$} where {\blue$\bar{x}_{2}=x_{2}+2.5x_{1}.$}
Also, we obtain $\bar{\gamma}_{Z_{2}}^{x_{3}}=2s$, $\bar{\gamma}_{X_{2}}^{x_{3}}=s$.
The bound of $\xi(t)$ can be calculated as follows
\[
|\xi(t)|=\left|\frac{d\vartheta_{2}(\bar{x}_{2})}{d\bar{x}_{2}}\dot{\bar{x}}_{2}\right|\leq\max\{\gamma_{\xi}^{z}(\|\h{z}_{2}\|),\gamma_{\xi}^{x}(\|\h{\bar{x}}_{2}\|),\gamma_{\xi}^{r}(\|r\|)\}
\]
with $\gamma_{\xi}^{z}(s)=2.5s^{2}+12.5s$, $\gamma_{\xi}^{x}(s)=0.27s^{5}+3.56s^{4}+15s^{3}+40s^{2}+70s$,
and $\gamma_{\xi}^{r}(s)=s^{2}+5s$. The calculation shows that $\gamma(s)$
in (\ref{eq:gamma-1}) is $\gamma(s)=\gamma_{\xi}^{x}(s)$ and it
satisfies $\lim_{s\rightarrow0^{+}}\gamma(s)/s<\infty.$  Let $\bar{\gamma}(s)$
in (\ref{eq:gamma-1}) be $\bar{\gamma}(s)=1/0.99\gamma(s)$. Then,
the event-triggered law (\ref{eq:t_bar-1}) can achieve the stabilization
without Zeno behavior. 
Figure \ref{fig:sim2} shows that
the sampling interval $\Delta_{k}=t_{k+1}-t_{k}$ converges to $0.0141$s
as $t\rightarrow\infty$, due to $\lim_{s\rightarrow0^{+}}\bar{\gamma}(s)/s=70.7$
(by Proposition \ref{prop:sp}) and no Zeno behavior occurs.
%The simulation and the comparison result of the proposed event-triggered law and periodic sampling  can be found in the more detailed version \cite{zhu2018stabilization}.
Figure \ref{fig:sim1} shows that the state
asymptotically goes to zero. 
 Note
that $z_{1}$-dynamics do not converge to zero exponentially and
 the triggering law using exponentially converging threshold signals
may lead to Zeno behavior (see Example 2 in \cite{Liu2015a}), while
the proposed controller does not. 
Next, let us compare the proposed event-triggered law and periodic sampling with sampling interval being $0.0141$s when the initial condition is $[z_1,x_1,z_2,x_2]\t=[-19.67,23.09,-24.77,13.755]\t$. 
The simulation in Figure \ref{fig:sim3} shows that  the proposed event-triggered law can stabilize the system while the periodic sampling  does not. This is consistent with result in \cite{Chen2014a} that the  sampling interval  required to stabilize the plant depends on  initial condition.

% \begin{figure}
% \begin{minipage}[t]{0.45\columnwidth}%
% \centering\includegraphics[scale=0.4]{state}
%
% (a)%
% \end{minipage}\qquad{} %
% \begin{minipage}[t]{0.45\columnwidth}%
% \centering\includegraphics[scale=0.4]{sp}
%
% (b)%
% \end{minipage}\caption{\label{fig:sim} Simulation profiles of the event-triggered control for
% the system (\ref{eq:xzxz}).}
% \end{figure}

\section{Conclusion \label{sec:con}}
%{\color{red}
%For the event-triggered control system, sometimes it is required to be robust to imperfections, such as external disturbances, modeling uncertainties, measurement noise, times delays, and so on. Some results concerning the robustness of the event-triggered control systems have been provided in \cite{brunner2019event,Liu2018Event,borgers2018periodic,khashooei2017output,dolk2017output,borgers2014event}. In particular, in , the robustness of the event-triggered mechanism with respect to time delays has been investigated, in  with respect to the state quantization, and in  with respect to the external disturbances. The robustness problem under the event-based control systems with both disturbances and uncertainties is discussed.
%}
In this paper, a novel event-triggered control scheme for nonlinear
systems was proposed. Then, it was applied to develop a theorem for
the event-triggered control of interconnected nonlinear systems. We
also provided an easy-to-check Zeno-free condition and showed that
the sampling interval converges to a constant as time approaches infinity.
Then, we employed the theorem to solve the event-triggered control
for lower-triangular systems with dynamic uncertainties. 
%A continuous-time
%stabilization controller was proposed followed by the design of the
%event-triggered law. Finally, the numerical simulation showed the
%effectiveness of the proposed method.  
{\color{blue}
 It remains an interesting future direction to consider the robustness of the proposed  with respect to more imperfections, such as  measurement noise 
in \cite{brunner2019event,dolk2017output,borgers2018periodic}, %borgers2014event ,khashooei2017output
state quantization in \cite{Liu2018Event}, modeling uncertainties  in \cite{Kishida2018Event}, time delays in \cite{borgers2018riccati}, and so on. }

\bibliographystyle{plain}
\bibliography{lit}

\end{document}